\documentclass[aps,nofootinbib,amsmath,prd,onecolumn,notitlepage,showpacs,superscriptaddress,groupedaddress,11pt]{revtex4-1}

\usepackage[american]{babel}
\usepackage{amsfonts}
\usepackage{amsmath}
\usepackage{amssymb}
\usepackage{slashed}
\usepackage{xcolor}
\usepackage{bm}
\usepackage{float}
\usepackage[utf8]{inputenc}
\usepackage[macrosonly]{chet}
\usepackage{nicefrac}
\usepackage{hyperref}
\usepackage{url}

\definecolor{darkblue}{rgb}{0.1,0.1,0.7}
\hypersetup{colorlinks,
           linkcolor={darkblue},
           citecolor={darkblue},
           urlcolor={darkblue}
}

\newcommand{\Diff}{{\rm \emph{Diff}}}
\newcommand{\Weyl}{{\rm \emph{Weyl}}}

\newcommand{\lvec}[2]{\raise #1\hbox{$^\leftarrow$} \hspace{-9pt} #2}
\newcommand{\rvec}[2]{\raise #1\hbox{$^\rightarrow$} \hspace{-9pt} #2}
\newcommand{\lrvec}[2]{\raise #1\hbox{$^\leftrightarrow$} \hspace{-9pt} #2}

\newcommand{\lsp}{\hspace{0.5pt}}

\allowdisplaybreaks

\usepackage[normalem]{ulem}

\allowdisplaybreaks

\begin{document}

\title{Weyl Covariance and the Energy Momentum Tensors \\
of Higher-Derivative Free Conformal Field Theories}

\author{Andreas Stergiou}
\email{andreas.stergiou@kcl.ac.uk}
\affiliation{Department of Mathematics, King's College London, Strand, London WC2R 2LS, United Kingdom}

\author{Gian Paolo Vacca}
\email{vacca@bo.infn.it}
\affiliation{INFN - Sezione di Bologna, via Irnerio 46, 40126 Bologna, Italy}

\author{Omar Zanusso}
\email{omar.zanusso@unipi.it}
\affiliation{
Universit\`a di Pisa and INFN - Sezione di Pisa, Largo Bruno Pontecorvo 3, 56127 Pisa, Italy}

\begin{abstract}
%
Energy momentum tensors of higher-derivative free scalar conformal field
theories in flat spacetime are discussed. Two algorithms for the
computation of energy momentum tensors are described, which accomplish
different goals: the first is brute-force and highlights the complexity of
the energy momentum tensors, while the second displays some features of
their geometric origin as variations of Weyl invariant curved-space
actions. New compact expressions for energy momentum tensors are given and
specific obstructions to defining them as conformal primary operators in
some spacetime dimensions are highlighted. Our discussion is also extended
to higher-derivative free spinor theories, which are based on
higher-derivative generalizations of the Dirac action and provide
interesting examples of conformal field theories in dimension higher than
two.
\end{abstract}

\pacs{}
\maketitle

\section{Introduction}\label{sect:intro}

The energy momentum tensor (EMT) is a fundamental part of the spectrum of
local conformal field theories (CFTs). It represents the Noether current
for spacetime translations, and it can be used to define currents
associated with symmetries of the conformal group. Two- and higher-point
functions of EMTs include important observables of CFTs that have a variety
of interesting properties and applications.

This work intends to contribute to the extensive body of work on EMTs by
deriving them for higher-derivative free theories and analyzing their
properties both from an algebraic and a geometric point of view. We focus
mostly on higher-derivative scalar theories, which have also been discussed
in~\cite{Osborn:2016bev, Guerrieri:2016whh, Brust:2016gjy}. These theories
constitute important toy models for conformal symmetry in higher
dimensions.  They are models in which the equations of motion involve an
arbitrary power of the Laplacian acting on a scalar field.
Higher-derivative free scalar theories are nonunitary CFTs which are simple
enough that we can obtain several results analytically, but nontrivial
enough that we can test modern methods and ideas of conformal symmetry and
geometry.

One important aspect of higher-derivative free theories is that we know how
to couple them to curved space in a Weyl invariant way.  This gives us the
opportunity to compute their energy momentum tensors (EMTs) using the
variational definition customary in general relativity and discuss their
properties in the flat space limit, where the EMTs become operators of the
corresponding CFT. It is crucial to stress that Weyl invariance implies
conformal invariance, i.e.\ the flat space restriction of a Weyl invariant
action gives rise to a conformal theory, but the converse is not true. This
happens because when we attempt to lift a CFT to curved space there may
exist obstructions to constructing Weyl covariant operators using curvature
tensors.  The discussion of these obstructions relates to important results
in the mathematical literature with salient ramifications in differential
geometry.

In general, the presence of obstructions to a Weyl invariant formulation
when lifting the CFT to curved space is signalled by nonzero contributions
to the trace of the EMT in lower even dimensions. Among these, the special
case is the continuation to two dimensions, in which the conformal group
admits the Virasoro extension. To give a concrete example, the simplest
nontrivial theory with equations of motion $\Box^2\phi=0$ is a CFT in
$d=2$, but has an obstruction to the Weyl invariant formulation in $d=2$.
As a consequence, the trace of the naive EMT is nonzero (it is not a
conformal primary), and the theory is invariant only under the subgroup
of global conformal transformations, instead of the full Virasoro group
\cite{Nakayama:2016dby}.

In this paper we discuss the computation of the EMTs of higher-derivative
theories starting from the Weyl invariant formulation in curved space.
This allows us to conjecture a general form for the corresponding EMTs,
interpreted as tensors of CFTs, in which there are still signatures of the
original Weyl invariance and, especially, of the obstructions to Weyl
invariance. Our conjectures are supported with explicit computations of
EMTs. For this purpose, we discuss two algorithms for the computation of
EMTs for the first few higher-derivative free theories.  The algorithms are
complementary in that the first allows us to compute the EMTs as primary
tensors of the corresponding CFTs but fix their normalization under the
assumption that they descend from a Weyl invariant theory, resulting in
expressions which manifestly show the obstructions to Weyl and conformal
invariance in lower even dimensions. The second one instead follows a
general improvement procedure, but allows us to rewrite the improvements in
terms of contributions, which are given by very compact expressions that
can be argued to originate from the terms in the Weyl invariant actions.

In our discussion we find that although higher-derivative free scalar
theories can generically be lifted to curved space in a Weyl invariant way,
in some even spacetime dimensions the existence of certain ``obstruction
tensors'' does not allow such a lift. In these problematic spacetime
dimensions, the obstruction tensors are responsible for our inability to
construct conformal primary EMTs. The lowest-dimension obstruction tensor
is the Bach tensor, which appears in the six-derivative theory and does not
allow the definition of a primary EMT for this theory in $d=4$. This was
first observed in~\cite{Osborn:2016bev}, and is here generalized to
higher-derivative theories. As we will see below, the $2n$ derivative free
theory with $n\geq3$ does not admit a conformal primary EMT for even
integer spacetime dimensions $d$ that satisfy $4\leq d\leq 2n-2$. This has
already been observed in the literature both in
mathematics~\cite{Graham-existence, Graham-nonexistence,
Gover-nonexistence} and in physics~\cite{Karananas:2015ioa, Osborn:2016bev,
Brust:2016gjy, Farnsworth:2017tbz}, but we provide a direct connection to
obstruction tensors in curved space and to defining the energy-momentum
tensor as a primary operator of the CFT.

We conclude the paper by extending several of the above ideas to
higher-derivative \emph{fermionic} theories, based on higher-derivative
generalizations of the massless Dirac equation.  We point at the existence
of a less known family of Weyl invariant higher-derivative Dirac spinors,
which also constitute CFTs in the flat space limit and dimension larger
than two. Although not analyzed in detail, we point to results in the
literature that show that obstruction tensors appear in this case as well.

This paper is organized as follows. In the next section we describe the
general setting for the scalar theories we are considering, propose a
general formula for the EMT of these theories that features specific
differential operators, and discuss its curved-space origin. We also
identify the curved-space source of the obstruction to defining the EMTs in
some even spacetime dimensions. In Sec.\ \ref{sect:ansatz} we explain and
use a brute-force algorithm for the construction of the EMTs. In
Sec.~\ref{sect:improvement-minimal} these EMTs are expressed in a more
compact form inspired by the curved space considerations of
Sec.~\ref{sect:weyl}. Unitarity bounds on the two- and three-point
functions of the EMT are examined in Sec.\ \ref{sect:unitarity}.  Free
Dirac fermion theories are considered in Sec.~\ref{sect:dirac} and we
conclude in Sec.~\ref{sect:conclusion}. Two appendices are included on the
action of the generator of special conformal transformations (Appendix
\ref{appendix:K}) and an explicit expression of the EMT of the scalar
theory with the ten-derivative kinetic term (Appendix \ref{appendix:d10}).

\section{From Weyl covariance to conformal invariance (and back)}\label{sect:weyl}

\subsection{Generalities}

Consider a $d$-dimensional Riemannian manifold equipped with a metric $g_{\mu\nu}$ and some internal vector bundle $V$.\footnote{%
The manifold can be either Riemaniann or pseudo-Riemannian. We tacitly assume
the latter with mostly-plus signature, but all conclusions apply to the former.
}
Let $\Delta$ be a linear covariant differential operator built with the metric $g_{\mu\nu}$
and its symmetric compatible connection $\nabla_\mu$ extended to $V$.
We choose $\Delta$ to be of rank $s$, implying that it is of the form
$\Delta \sim \partial^s+\cdots$,
where in the dots we hide terms with less than $s$
derivatives and curvatures.
We take the operator to have special covariance properties under Weyl rescalings:
for a transformation of the metric of the form
$g_{\mu\nu} \to g'_{\mu\nu}= e^{2 \sigma}g_{\mu\nu}$ and $\sigma=\sigma(x)$,
the operator transforms as
\begin{equation} \label{eq:general-transf}
 \begin{split}
  \Delta & \to \Delta' = e^{-\frac{d+s}{2} \sigma} \Delta e^{\frac{d-s}{2} \sigma}\,.
 \end{split}
\end{equation}
In this case, the operator $\Delta$ is said to be Weyl covariant (or conformally covariant) with a bi-degree related to the numbers $s$ and $d$ \cite{Erdmenger:1997gy,Erdmenger:1997wy}.

Using $\Delta$ and \eqref{eq:general-transf}, we can construct
a Weyl invariant action. Introducing a field $\Phi$ on $V$ of dimension $\frac{d-s}{2}$,
which thus transforms as
$\Phi \to \Phi' = e^{-\frac{d-s}{2} \sigma}\Phi$
and its conjugate $\overline{\Phi}$ of the same dimension,
we have that the action
\begin{equation}\label{eq:general-action}
 \begin{split}
   S[\Phi] & = -\int d^{\lsp d}x \sqrt{-g}\, \overline{\Phi} \Delta \Phi
 \end{split}
\end{equation}
is invariant under Weyl transformations in an arbitrary dimension by construction. In general,
there can be geometrical obstructions to the construction of operators
with these properties for specific values of $d$ and internal indices of $\Phi$.
Nevertheless, there exist interesting families
of operators that satisfy \eqref{eq:general-transf}.

The most famous family includes the Graham-Jenne-Mason-Sparling
(GJMS) rank-$2n$ operators
$\Delta_{2n}$ for $n\in \mathbb{N}_0$ \cite{Graham-existence,Graham-nonexistence,Gover-nonexistence}, which act on real scalar fields $\phi$ of dimension $\delta_{2n}=\frac{d-2n}{2}$. They can be written as
\begin{equation}\label{eq:delta-operators-split}
 \begin{split}
  \Delta_{2n} &= \Delta_{2n,1} + \frac{d-2n}{2} Q_{2n}\,,
 \end{split}
\end{equation}
where the two terms separate the purely derivative part $\Delta_{2n,1}$
and the curvature part. By construction $\Delta_{2n,1}\phi_0=0$ for some constant field $\phi_0$. The entities $Q_{2n}$ are Branson's $Q$-curvatures \cite{Branson_1985}; if integrated
they provide
a generalization of Euler's characteristic in arbitrary dimension $d$ up to equivalence classes of conformally related metrics.
The importance of $Q$-curvatures in our discussion is highlighted in Sec.~\ref{sect:interpretation}.
The first member of this family is the well-known Yamabe operator for $n=1$
\begin{equation}\label{eq:Delta2}
 \begin{split}
  & \Delta_{2} = -\nabla^2 + \frac{d-2}{4(d-1)} R\,, \qquad
  \Delta_{2,1} = -\nabla^2\,, \qquad
  Q_2 = \frac{R}{2(d-1)}\,.
 \end{split}
\end{equation}
The pole of $\Delta_2$ for $d=1$ highlights an obstruction of extending Weyl invariance to a one-dimensional manifold due to the triviality of the geometry
in this case \cite{Brust:2016gjy}. In general the operator $\Delta_{2n}$
has poles for $d=1$ and $d=2,\ldots,2n-2$ (even) highlighting similar obstructions
based on the absence of Weyl covariant curvature tensors in lower-dimensional
geometries \cite{Farnsworth:2017tbz}.
Going back to $d=2$ dimensions and $n=1$, there is only one conformal class of metrics for each topology, so $Q_2$
itself coincides with Euler's density modulo a normalization, but this is not the case for $d=2n$ and $n\geq 2$, in which case there are more conformal classes on general manifolds.
The explicit form of the operators $\Delta_{2n}$ is known up to $n=4$ (eight derivatives), while implicit formulas are available for any $n$. The complexity of the $\Delta_{2n}$ operators increases with $n$ very quickly.

Following the logic that leads
to the construction of \eqref{eq:general-action},
each operator $\Delta_{2n}$ returns a Weyl invariant quadratic action for a scalar field
of dimension $\delta_{2n}=\frac{d-2n}{2}$ of the form
\begin{equation}\label{eq:general-action-sc}
 \begin{split}
   S[\phi,g] & = -\frac{1}{2}\int d^{\lsp d}x \sqrt{-g}\, \phi \Delta_{2n}\phi \,,
 \end{split}
\end{equation}
which we now normalize
with the factor $\frac{1}{2}$ to account for the reality of
the scalar field $\phi$. We have also made explicit the dependence of the action on the  metric. By construction, \eqref{eq:general-action-sc} is invariant under both Weyl transformations, which infinitesimally are
\begin{equation}\label{eq:weyl-transf}
 \begin{split}
 \delta^W_\sigma g_{\mu\nu} = 2\sigma g_{\mu\nu}
 \,, \qquad
 \delta^W_\sigma\phi = - \delta_{2n} \sigma \phi\,,
 \end{split}
\end{equation}
and reparametrizations, which are the diffeomorphisms
\begin{equation}\label{eq:diff-transf}
 \begin{split}
 \delta^E_\xi g_{\mu\nu} = \nabla_\mu\xi_\nu + \nabla_\nu\xi_\mu
 \,, \qquad
 \delta^E_\xi\phi = \xi^\mu \partial_\mu \phi\,.
 \end{split}
\end{equation}
When combined, the transformations form the group
$\Diff \ltimes \!\!{\Weyl}$.
The equations of motion for the field $\phi$
of \eqref{eq:general-action-sc} are simply
$\Delta_{2n}\phi=0$.

Weyl invariance is known to imply conformal invariance when explicitly choosing
a fixed ``background'' metric manifold.
For example, in the flat space limit, we have that
\begin{equation}
 \begin{split}
  \left.\Delta_{2n}\phi\right|_{\rm flat} =(-\partial^2)^n \phi=0\,,
 \end{split}
\end{equation}
which are known to be the equations of motion of (higher-derivative free) conformal field theories for any $n$ in any $d$ \cite{Brust:2016gjy}. This includes also the dimensionalities
$d$ for which there are obstructions
to Weyl invariance \cite{graham2005ambient},
which is a manifestation of the general fact that Weyl invariance implies conformal invariance, but the converse is not true \cite{Karananas:2015ioa}. In general, by selecting any specific Riemaniann manifold with metric $\overline{g}_{\mu\nu}$ in \eqref{eq:general-action-sc}, it is easy to see that the
resulting theory is invariant under the combination of diffeomorphisms and Weyl transformations that leave $\overline{g}_{\mu\nu}$ invariant,
$ \delta^W_\sigma \overline{g}_{\mu\nu}+\delta^E_\xi \overline{g}_{\mu\nu}=0$,
which generates the group of conformal isometries
on $\overline{g}_{\mu\nu}$.
Using the definition of conformal isometry, the conformal group is generated by conformal Killing vectors such that
$\overline{\nabla}_\mu\xi_\nu+\overline{\nabla}_\nu\xi_\mu
=\frac{2}{d}\overline{\nabla}\cdot \xi \, \overline{g}_{\mu\nu}$
in $d>2$.\footnote{%
A simple derivation. By taking the trace of
$\delta^W_\sigma \overline{g}_{\mu\nu}+\delta^E_\xi \overline{g}_{\mu\nu}= 2 \sigma \overline{g}_{\mu\nu} + \overline{\nabla}_\mu\xi_\nu+\overline{\nabla}_\nu\xi_\mu=0$,
it is straightforward to find the relation
$\sigma=-\frac{1}{d} \overline{\nabla}\cdot \xi$
between the infinitesimal parameters of the transformations
(indices are lowered with the metric, here $\overline{g}_{\mu\nu}$, so $\xi_\mu=\overline{g}_{\mu\nu} \xi^\nu$).
Substituting $\sigma$ in the original equation the result is the
equation satisfied by conformal Killing vectors, $\overline{\nabla}_\mu\xi_\nu+\overline{\nabla}_\nu\xi_\mu
-\frac{2}{d}\overline{\nabla}\cdot \xi \, \overline{g}_{\mu\nu}=0$.
The infinitesimal conformal isometry is defined
in general by the combination
$\delta^C_\xi = \delta^E_\xi + \delta^W_\sigma$
with $\sigma$ satisfying the above relation.
The metric $\overline{g}_{\mu\nu}$ is invariant by definition,
$\delta^C_\xi \overline{g}_{\mu\nu}=0$,
assuming that the algebra of conformal Killing vectors is nontrivial
(i.e.~there are nontrivial solutions to the conformal Killing equation,
otherwise the group is trivial and the only solution is $\xi^\mu=0$).
}

Notice that the conformal group might be trivial for a general metric manifold. For this reason, Weyl invariance is actually relevant for conformal invariance when the background metrics admit nontrivial solutions of the Killing equation. This is true, for example, when the background metrics are conformally flat, which we take to be the case given that we are ultimately interested in the flat space limit. 
The conformal group acts on $\phi$ as
\begin{equation}
 \begin{split}
\delta^C_\xi \phi= \xi^\mu \overline{\nabla}_\mu \phi +\frac{\delta_{2n}}{d} \overline{\nabla}\cdot \xi \phi\,.
 \end{split}
\end{equation}
In flat Minkowski space $\overline{g}_{\mu\nu}=\eta_{\mu\nu}$, the
conformal group is generated by the Poincar\'e group (translations,
rotations and boosts), as well as dilatations $x^\mu \to x^{\prime \mu}=
\lambda x^\mu$ and special conformal transformations. These transformations
cause specific rescalings of the metric in curved space, i.e.\
$g_{\mu\nu}\to e^{2\sigma(x)}g_{\mu\nu}$ for specific $\sigma(x)$, but Weyl
transformations in curved space allow for a completely general $\sigma(x)$.
Thus, the specific rescaling transformations that arise from conformal
transformations form a subgroup of the $\Diff \ltimes \!\!{\Weyl}$ group.

Going back to curved space, under an arbitrary transformation of both the field and the metric, the action \eqref{eq:general-action-sc} transforms as
\begin{equation}\label{eq:total-diff-sc}
 \begin{split}
   \delta S &= -\int d^{\lsp d}x \sqrt{-g} \Bigl( \delta\phi \Delta_{2n} \phi
  - \frac{1}{2} T^{\mu\nu} \delta g_{\mu\nu}\Bigr) \,,
 \end{split}
\end{equation}
where we defined the variational energy-momentum tensor (EMT) off-shell
\begin{equation}\label{eq:belifante-def}
 \begin{split}
  T^{\mu\nu} &= \frac{2}{\sqrt{-g}}\frac{\delta S}{\delta g_{\mu\nu}} \,.
 \end{split}
\end{equation}
Using the invariance properties of $S$ under infinitesimal
reparametrizations \eqref{eq:diff-transf}
and Weyl transformations \eqref{eq:weyl-transf},
it is straightforward to see that the variational EMT satisfies the off-shell relations
\begin{equation}\label{eq:belifante-eqns}
 \begin{split}
  \nabla_\mu T^{\mu\nu} = -\nabla^\nu \phi \Delta_{2n} \phi\,, \qquad
  g_{\mu\nu} T^{\mu\nu} = -\delta_{2n} \phi \Delta_{2n} \phi \,.
 \end{split}
\end{equation}
The immediate consequence is that the variational EMT is traceless
and conserved when evaluated on-shell using the equations of motion of $\phi$
\begin{equation}\label{eq:belifante-cons}
 \begin{split}
  \left.\nabla_\mu T^{\mu\nu}\right|_{\rm on-shell}= \left. g_{\mu\nu} T^{\mu\nu}\right|_{\rm on-shell}=0\,.
 \end{split}
\end{equation}
As a result,
if we take the limit $g_{\mu\nu}=\eta_{\mu\nu}$, we have that
$T^{\mu\nu}$ is the (improved) stress energy tensors of the corresponding
CFTs, which we know already to be the higher-derivative
free CFTs satisfying $(-\partial^2)^n \phi=0$ in flat space.
Using \eqref{eq:belifante-cons}, we also know that off-shell in flat space\footnote{These relations for general $n$ were conjectured in~\cite{Osborn:2016bev} and used in~\cite{Safari:2021ocb}.}
\begin{equation}\label{eq:belifante-cons-flat}
 \begin{split}
  \partial_\mu T^{\mu\nu}|_{\rm flat}=
  -\partial^\nu \phi (-\partial^2)^n \phi\,,
  \qquad
  T^{\mu}{}_{\mu}|_{\rm flat}=
  -\delta_{2n} \phi (-\partial^2)^n \phi
  \,.
 \end{split}
\end{equation}
As a consequence, we can institute a one-to-one correspondence between the Weyl covariant GJMS operators $\Delta_{2n}$ in general $d$
and the EMT of higher-derivative free CFTs on the basis of the variational definition \eqref{eq:belifante-def} of $T^{\mu\nu}$
obtained from a conformally flat geometry.

\subsection{The improved primary EMT of flat space CFT}

Now we restrict again our attention to flat space.  With $T^{\mu\nu}$ being
the EMT of a CFT, it is natural to ask if it is also a primary operator,
that is, if it satisfies $\hat{K}_\mu T^{\rho\theta}=0$ at the origin
$x^\mu=0$ of $\mathbb{R}^d$, having introduced $\hat{K}_\mu$ as the
generator of special conformal transformations in flat space. Given the
explicit forms of the operator $\Delta_{2n}$ for $n=1,2,3$, it is possible
to check directly that $\hat{K}_\mu T^{\rho\theta}=0$ for these cases,
implying that the variational formula not only gives the traceless
conserved EMT in flat space, but that such EMT is also a primary operator
in the usual sense of CFT \cite{Osborn:2016bev}.  A primary EMT for $n=4$
was first given in~\cite{Guerrieri:2016whh}.  An algorithm for the
computation of the action of $\hat{K}_\mu$ is given in
Appendix~\ref{appendix:K}.

To explore the structure of the primary EMT further, suppose that we start
from a non-improved EMT, denoted $T_c^{\mu\nu}$, which is conserved,
$\partial_\mu T_c^{\mu\nu}=0$, but not (necessarily) traceless,
$\delta_{\mu\nu}T_c^{\mu\nu} \neq 0$, for the CFT with
$(-\partial^2)^n\phi=0$. The origin of $T_c^{\mu\nu}$ is not particularly
important, although later on we will make a specific choice. One may
assume, for example, that it could have been derived from the condition
$\partial_\mu {\cal L}=0$, where ${\cal L}={\cal
L}(\phi,\partial\phi,\cdots)$ is a Lagrangian for the higher-derivative
free theory. We also assume, for simplicity, that $T_c^{\mu\nu}$ is
symmetric, although what follows can be discussed equally well starting
from a nonsymmetric tensor \cite{OsbornLectures}.

Given that the theory with equations of motion $(-\partial^2)^n\phi=0$ is a CFT with a proper EMT in $d\geq 2n$,
we know that, even though $T_c^{\mu\nu}$ is not traceless, it is possible to find on-shell a symmetric tensor $Z^{\mu\nu}$ for which
\begin{equation}\label{eq:Z-def}
 \begin{split}
 \delta_{\mu\nu}T_c^{\mu\nu}= \partial_\mu\partial_\nu Z^{\mu\nu}\,.
 \end{split}
\end{equation}
As a consequence, we can define an improved EMT $T_i^{\mu\nu}$ as
\begin{equation}\label{eq:improved-emt-0}
 \begin{split}
 T_i^{\mu\nu}
 =T_{c}^{\mu\nu}+{\cal D}^{\mu\nu}{}_{\alpha\beta} Z^{\alpha\beta}\,,
 \end{split}
\end{equation}
where we introduced the differential operator ${\cal D} \sim \partial^2$,
which is the unique operator constructed with two partial derivatives
and properties $ \partial_\mu {\cal D}^{\mu\nu\rho\theta}=0$
and
$\delta_{\mu\nu}{\cal D}^{\mu\nu\rho\theta}=-\partial^\rho\partial^\theta$
(the explicit form is given below).
One can check easily that $T_i^{\mu\nu}$ is both conserved and traceless,
so it is the ``correct'' EMT for the CFT.

However, $T_i^{\mu\nu}$ is not necessarily a primary operator,
or at least it is not
for an arbitrary choice of $Z^{\mu\nu}$ making the EMT traceless.
This happens because \eqref{eq:Z-def} does not completely fix the form of
$Z^{\mu\nu}$, which has to be integrated and is subject to the equations of motion.
In general, there are multiple
options for $Z^{\mu\nu}$ that satisfy \eqref{eq:Z-def} in agreement
with $(-\partial^2)^n\phi=0$ (this is
particularly relevant for $n\geq 3$, as we discuss later).
Another point of view on this
matter comes from noticing that
\begin{equation}\label{eq:improved-emt}
 \begin{split}
 T_i^{\prime\mu\nu}
 =T_{c}^{\mu\nu}+{\cal D}^{\mu\nu}{}_{\alpha\beta} Z^{\alpha\beta}
 +\lambda {\cal D}_B^{\mu\nu}{}_{\alpha\beta} \tilde{Z}^{\alpha\beta}\,,
 \end{split}
\end{equation}
is an equally ``correct'' traceless conserved EMT, where we have introduced an arbitrary
real parameter $\lambda$ and a new differential operator ${\cal D}_B \sim \partial^4$. The latter is the unique differential operator satisfying
$\delta_{\mu\nu}{\cal D}_B^{\mu\nu\rho\theta}=\delta_{\rho\theta}{\cal D}_B^{\mu\nu\rho\theta}=0$, $\partial_\mu{\cal D}_B^{\mu\nu\rho\theta}=0$ and, importantly, ${\cal D}_B^{\mu\nu\rho\theta}(\partial_\rho \xi_\theta+\partial_\theta \xi_\rho)=0$ for any vector $\xi^\mu$
(the explicit form is also given below).
In this situation, one can assume that, even if $T_i^{\mu\nu}$ is not primary, $T_i^{\prime\mu\nu}$ can be made
a primary operator by acting with $\hat{K}_\mu$ and checking that the
result is zero for some fixed value of the parameter $\lambda$. This should
be true at least for all the dimensionalities $d \geq 2n$ which allow a
proper primary EMT for the higher-derivative free CFTs.
This argument suggests that there is a unique primary traceless conserved EMT for the CFTs with equations of motion $(-\partial^2)^n\phi=0$,
which we choose to denote $T^{\mu\nu}$.

One crucial point in the above argument is the uniqueness of the differential operator ${\cal D}_B \sim \partial^4$ and the freedom that it allows in displacing the action of ${\cal D} \sim \partial^2$ on $Z_{\mu\nu}$.
As promised, the explicit form of the two operators~\cite{Osborn:2016bev} is
\begin{align}\label{eq:D-tensors}
 {\cal D}^{\mu\nu\rho\theta} &=
 \frac{1}{d-2}(\eta^{\mu(\rho} \partial^{\theta)} \partial^{\nu}
 +\eta^{\nu(\rho} \partial^{\theta)} \partial^{\mu}
 -\eta^{\mu(\rho} \eta^{\theta)\nu} \partial^2
 -\eta^{\mu\nu}\partial^\rho\partial^\theta)
 -\frac{1}{(d-1)(d-2)}\eta^{\rho\theta}(\partial^\mu\partial^\nu -\eta^{\mu\nu}\partial^2)\,,
 \nonumber\\
 {\cal D}_B^{\mu\nu\rho\theta} &={\cal D}^{\mu\nu\rho\theta}\partial^2
 -\frac{1}{d-1}(\partial^\mu\partial^\nu -\eta^{\mu\nu}\partial^2 )\partial^\rho\partial^\theta\,,
\end{align}
from which one can verify that they satisfy the required identities.
Modulo an overall normalization, ${\cal D}$ is truly unique on the basis of the requirements that it is constructed from the flat space metric and partial derivatives. On the other hand, ${\cal D}_B$ is not completely fixed
by the requirements above, not even including ${\cal D}_B^{\mu\nu\rho\theta}(\partial_\rho \xi_\theta+\partial_\theta \xi_\rho)=0$.
In fact, we could construct ${\cal D}_{B,m}\sim \partial^{4+2m}$ with the same properties, but more derivatives ($m\geq 1$).
Nevertheless, it is not difficult to check that ${\cal D}_{B,m}= (-\partial^2)^m {\cal D}_{B}$, always modulo normalization, which happens because ${\cal D}_{B,m}$ acts on a symmetric tensor and returns a symmetric tensor, so there is no room for further uncontracted derivatives that substantially change the tensor structures of \eqref{eq:D-tensors}.
As a consequence, there are no further possible improvements to \eqref{eq:improved-emt}
given that we can move $(-\partial^2)^m$ to the right and consequently reabsorb
any further tensor playing the same role of $\tilde{Z}_{\mu\nu}$ to match \eqref{eq:improved-emt}. A geometric interpretation of the uniqueness of ${\cal D}_{B}$ is postponed to Sec.~\ref{sect:interpretation}.

We are thus led to believe that \eqref{eq:improved-emt}, for a fixed value
of $\lambda$, is the general form of the primary EMT of the
higher-derivative free scalar theories.
The structure of \eqref{eq:improved-emt} clearly suggests a geometrical origin
of the primary EMT, which will be clarified in the next subsection.
To clarify one important point, it is always possible to rewrite
$T_i^{\prime\mu\nu}$ of \eqref{eq:improved-emt}
in the form \eqref{eq:improved-emt-0} for an opportune tensor $Z^{\mu\nu}$.
This can be shown easily using the explicit form of ${\cal D}_B$ as a function of ${\cal D}$ in \eqref{eq:D-tensors} and the fact that ${\cal D}_{\mu\nu}{}^{\rho\theta}(g_{\rho\theta} {\cal O})= {\cal P}_{\mu\nu}{\cal O}$,
for any scalar ${\cal O}$ and
\begin{equation}\label{eq:P-def}
 \begin{split}
{\cal P}_{\mu\nu}=(d-1)^{-1}(\partial_\mu\partial_\nu-\delta_{\mu\nu}\partial^2)\,.
 \end{split}
\end{equation}
The differential operator ${\cal P}$ comes about as the original improvement to an EMT, which is needed for the simplest cases with two derivatives.
Altogether, the operators satisfy the relations
\begin{equation}\label{eq:P-algebra}
 \begin{split}
 {\cal D}^{\mu\nu}{}_{\rho\theta} {\cal P}^{\rho\theta} &={\cal D}_B^{\mu\nu}{}_{\rho\theta} {\cal P}^{\rho\theta}=0 \,,
 \end{split}
\end{equation}
and
\begin{equation}\label{eq:D-algebra}
 \begin{split}
 &{\cal D}^{\mu\nu}{}_{\rho\theta} {\cal D}^{\rho\theta}{}_{\alpha\beta} = -\frac{1}{d-2}{\cal D}_B^{\mu\nu}{}_{\alpha\beta}
 \,, \qquad
 {\cal D}_B^{\mu\nu}{}_{\rho\theta} {\cal D}_B^{\rho\theta}{}_{\alpha\beta} = -\frac{1}{d-2}\partial^2\partial^2 {\cal D}_B^{\mu\nu}{}_{\alpha\beta}
 \\
 &{\cal D}^{\mu\nu}{}_{\rho\theta} {\cal D}_B^{\rho\theta}{}_{\alpha\beta} = {\cal D}_B^{\mu\nu}{}_{\rho\theta} {\cal D}^{\rho\theta}{}_{\alpha\beta}=-\frac{1}{d-2}\partial^2 {\cal D}_B^{\mu\nu}{}_{\alpha\beta} \,.
 \end{split}
\end{equation}
In order to find the $\lambda$-dependence
in a form such as \eqref{eq:improved-emt},
it is thus necessary to implement a form of
``naive'' improvement and isolate the operator $\tilde{Z}_{\mu\nu}$ that is responsible for the correction that makes the EMT a primary.
The forms of the operators $Z_{\mu\nu}$ and $\tilde{Z}_{\mu\nu}$ in \eqref{eq:improved-emt} are somewhat ambiguous, because the differential operators
${\cal D}$ and ${\cal D}_B$ admit, in general, a nontrivial kernel, corresponding to
zero modes under the on-shell condition $(-\partial^2)^n\phi=0$. The kernel can be used to simplify the final forms as we shall see later.

With all the above considerations in mind, we postulate a general form for the primary EMT of the CFT
\begin{equation}\label{eq:T-general-structure}
 T^{\mu\nu} = T_{c}^{\mu\nu}
 -{\cal P}^{\mu\nu} {\cal O}
 + {\cal D}^{\mu\nu}{}_{\rho\theta}{\cal S}^{\rho\theta}
 + {\cal D}_B^{\mu\nu}{}_{\rho\theta}\tilde{\cal S}^{\rho\theta}\,.
\end{equation}
If $T^{\mu\nu}$ is a tensor constructed from $\phi$ and $\delta_{\mu\nu}$
that contains $2n$ derivatives, we infer that ${\cal O}$ is a scalar with
$2n-2$ derivatives, ${\cal S}^{\mu\nu}$ is a symmetric tensor with $2n-2$ derivatives and no metric, and, finally, $\tilde{\cal S}^{\mu\nu}$ is a symmetric tensor with $2n-4$ derivatives and no metric. Consequently, ${\cal S}^{\mu\nu}$ is expected to contribute starting
from $n\geq 2$ (for $n=1$ it must be proportional to the metric, and thus it would give a scalar contribution), while, for similar reasons, $\tilde{\cal S}^{\mu\nu}$ is expected to contribute starting from $n\geq 3$. This pattern is verified below in the explicit computations.

We have now come full circle: on the one hand we have a family of unique
variational EMTs that come from varying with respect to the metric the Weyl
invariant actions \eqref{eq:general-action-sc} as in \eqref{eq:belifante-def}
and evaluating the result in flat space, which can be argued to belong to a
CFT. On the other hand, we have a family of primary EMTs of the
higher-derivative free models in flat space, which is presumably unique thanks to the fact that they are primary operators. The natural question is whether these two families are actually the same for any $n$, and we believe the answer affirmative, which is implicitly evident from the fact that we denoted the two EMTs with the same symbol. In the following, we discuss two rather efficient algorithms to compute $T^{\mu\nu}$ using information from either side of this question, inferring the uniqueness of the resulting
$T^{\mu\nu}$ and the structure \eqref{eq:T-general-structure} in relation to the differential operators given in \eqref{eq:D-tensors}.

\subsection{Geometrical interpretation -- the obstructions to Weyl invariance}
\label{sect:interpretation}

Let us try to frame the discussion that has lead to the form
\eqref{eq:T-general-structure} from a more geometrical standpoint in
relation to Weyl invariance and its obstructions. The operators
$\mathcal{P}^{\mu\nu}$ and $\mathcal{D}^{\mu\nu\rho\theta}$ arise from a
metric variation of terms in the Weyl invariant curved space action that
involve the Ricci scalar and the Ricci tensor, respectively. The remaining
tensor, $\mathcal{D}^{\mu\nu\rho\theta}_B$, arises from contributions that
involve the Bach tensor.

The Bach tensor is a rank-two, symmetric, traceless tensor with four
derivatives of the metric that is defined in $d\geq4$. It is additionally
Weyl covariant and divergenceless when $d=4$. In general $d\geq4$ it is
given by
\eqn{B_{\mu\nu}=\nabla^\lambda
C_{\mu\nu\lambda}-P^{\lambda\rho}W_{\lambda\mu\nu\rho}\,,}[]
where $P_{\mu\nu}=\frac{1}{d-2}(R_{\mu\nu}-g_{\mu\nu}\hat{R})$ with
$\hat{R}=\frac{1}{2(d-1)}R$ is the Schouten tensor,
$C_{\mu\nu\lambda}=\nabla_\lambda P_{\mu\nu}-\nabla_\nu P_{\mu\lambda}$ the
Cotton tensor and $W_{\lambda\mu\nu\rho}$ the Weyl tensor.  Specifically in
$d=4$, the Bach tensor arises as
\begin{equation} \label{eq:B-def}
 B^{\mu\nu}=-\frac14\frac{1}{\sqrt{-g}}
  \frac{\delta}{\delta g_{\mu\nu}}\int d^{\lsp 4}x\,\sqrt{-g}\,
W^{\kappa\lambda\rho\sigma}W_{\kappa\lambda\rho\sigma}\,,
\end{equation}
where the Weyl tensor is now defined in $d=4$. The Bach tensor is the first
member of a family of tensors with similar properties defined in even
$d\geq4$.

To see how \eqref{eq:B-def} can be generalized to an infinite family, we
notice first that in $d=4$ the $Q$-curvature ($Q_4$, which is defined
implicitly in \eqref{eq:delta-operators-split}) is, up to a total
derivative $\nabla^2\hat{R}$ term, a linear combination of
$W^{\mu\nu\rho\sigma}W_{\mu\nu\rho\sigma}$ and the Euler density
$E_4$.\footnote{Some useful expressions can be found in~\cite[Appendix
A]{Osborn:2015rna}. In general $d$ we have
$E_4=W^{\mu\nu\rho\sigma}W_{\mu\nu\rho\sigma}-4(d-2)(d-3)(P^{\mu\nu}P_{\mu\nu}-\hat{R}^2)$
and $Q_4=-\frac{1}{2(d-2)(d-3)}(W^{\mu\nu\rho\sigma}W_{\mu\nu\rho\sigma} -
E_4)+\frac{d-4}{2}\hat{R}^2-\nabla^2\hat{R}$. Crucially, the $\hat{R}^2$
term in $Q_4$ drops out in $d=4$.} The Euler density is a topological
invariant, so we can replace
$W^{\kappa\lambda\rho\sigma}W_{\kappa\lambda\rho\sigma}$ with $Q_4$ in
\eqref{eq:B-def}.  In other words, $\frac{1}{\sqrt{-g}}
\frac{\delta}{\delta g_{\mu\nu}} \int d^{\lsp 4}x \sqrt{-g}\, Q_4$ gives
the Bach tensor in $d=4$.  The above property generalizes to higher $d$,
i.e.~the $Q$-curvature in dimension $d$ is a linear combination of a
conformally invariant quantity and the $d$-dimensional Euler
density up to total derivatives~\cite{graham2005ambient}.\footnote{%
Another way to say this is that the $Q$-curvatures for $d=2n$ are topological
invariants up to a given conformal class of metrics.}
This means that for $n\geq2$ we can define a family of
tensors as the variational derivative of Branson's $Q$-curvature~\cite{graham2005ambient}
\begin{equation} \label{obtens}
\mathcal{O}_{2n}^{\mu\nu}=\frac{1}{\sqrt{-g}}
\frac{\delta}{\delta g_{\mu\nu}}\int d^{\lsp 2n}x\,\sqrt{-g}\, Q_{2n}\,,
\end{equation}
and obviously $\mathcal{O}_4^{\mu\nu} = B^{\mu\nu}$. (Notice that
individually each $Q$-curvature depends on $d$, so we are taking $d=2n$
for each one of them.)
These tensors are called
\emph{obstruction tensors} and were first discussed in~\cite{FG1}.
Crucially for us, the tensors $\mathcal{O}_{2n}^{\mu\nu}$ take the form
\begin{equation} \label{obsform}
  \mathcal{O}_{2n}^{\mu\nu}\sim(\nabla^2)^{n-2}
B^{\mu\nu}+\text{(terms with higher powers of the curvature)}\,,
\end{equation}
where $B^{\mu\nu}$ is the Bach tensor defined in dimension $d=2n$.  To
obtain the EMT from curved space as in \eqref{eq:belifante-def}, we need to
take one functional derivative of the appropriate action with respect to
the metric and then restrict the result to the flat metric.  In our case
the appropriate action for the $2n$-derivative theory with $n\geq3$ will
contain the obstruction tensor $\mathcal{O}_{2n-2}^{\mu\nu}$ and it is only
the first term on the right-hand side of \eqref{obsform} that contributes
to the EMT in the limit.

The variation of the Bach tensor is precisely what gives rise to the
differential operator $\mathcal{D}_B$ in \eqref{eq:T-general-structure} as
discussed in Ref.~\cite{Osborn:2016bev} for the case $d=6$.  From direct
inspection of \eqref{obsform}, we then expect that, in arbitrary positive
even $d$, further obstructions can appear only in the form $\mathcal{D}_B$
and be, at most, multiplied by some power of the flat space Laplacian. The
uniqueness of $\mathcal{D}_B$ modulo powers of the flat space Laplacian is
precisely what we observed in the discussion enveloping the definition
\eqref{eq:D-tensors}.  As we will see explicitly below, obstruction tensors
provide an obstruction to being able to define a primary $T^{\mu\nu}$ for
the $2n$-derivative theory with $n\geq3$ when $d$ is even and satisfies
$4\leq d\leq 2n-2$.

\section{EMT from ansatze}\label{sect:ansatz}

Now we focus entirely on the higher-derivative free theories with
equations of motion $(-\partial^2)^n\phi=0$ in flat space. We can provide a general parametric ansatz for $T^{\mu\nu}$ which is the sum of all possible
monomials constructed with two copies of $\phi$, $n$ derivatives $\partial^\mu$,
and the metric $\eta^{\mu\nu}$, weighted by opportune coefficients
that should be determined by the requirements of it being the EMT of a CFT.

The four conditions that the primary EMT of a CFT should satisfy are that it
is symmetric, conserved, traceless and primary in $d>2$. In formulas
\begin{equation}\label{eq:main-conditions}
 \begin{split}
 T^{[\mu\nu]} |_{\rm on-shell} =
 \partial_\mu T^{\mu\nu}|_{\rm on-shell} =
 g_{\mu\nu} T^{\mu\nu}|_{\rm on-shell} =
 \hat{K}_\rho T^{\mu\nu}(0)|_{\rm on-shell} =0\,.
 \end{split}
\end{equation}
The application of $\hat{K}_\rho$ at $x^\mu=0$ can be carried out using the
algorithm presented in \cite{Osborn:2016bev} and briefly summarized in the appendix.
The tensors of a CFT are fixed up to an overall normalization, so we do not expect these conditions to completely fix the structure of $T^{\mu\nu}$ neither on- nor off-shell. The off-shell form of $T^{\mu\nu}$ can still depend parametrically on a contribution of the form $\eta^{\mu\nu}\phi \partial^{2n} \phi$ that goes to zero on-shell, as well as on the overall normalization.

To completely fix all parametric dependencies of $T^{\mu\nu}$, we require that
the EMT should comply with the general off-shell form that comes
from varying the Weyl invariant action \eqref{eq:belifante-cons-flat}, which in flat space are
\begin{equation}\label{eq:extra-conditions}
 \begin{split}
 \left.\partial_\mu T^{\mu\nu}\right|_{\rm off-shell} = - \partial^\nu \phi (-\partial^2)^n \phi\,,
 \qquad
 \left. g_{\mu\nu} T^{\mu\nu}\right|_{\rm off-shell} =-  \frac{d-2n}{2} \phi (-\partial^2)^n \phi
 \,.
 \end{split}
\end{equation}

In other words, the divergence of $T^{\mu\nu}$ goes to zero as a consequence of the equation of motion going to zero (and not their derivatives, e.g.~there is no term $\phi \partial^\nu (-\partial^2)^2\phi$ in the divergence) and the trace relates to the scaling dimension of $\phi$.
We have checked explicitly that for $n=1,2,3$ the combined requirements
\eqref{eq:main-conditions} and \eqref{eq:extra-conditions} return a unique
parameter-independent $T^{\mu\nu}$ which matches the variational approach
of the Weyl covariant action \eqref{eq:general-action-sc}.
We have also computed the unique primary EMTs up to $n=8$ (sixteen derivatives).

\subsection{Example of the procedure for \texorpdfstring{$n=1$}{n=1}}

To highlight the procedure, consider the simplest theory with two derivatives and equations of motion $\partial^2\phi=0$. The most general ansatz for the stress-energy tensor contains all possible terms constructed with a metric, two derivatives and two copies of the field $\phi$.
We find that there are only four independent monomials
\begin{equation}\label{eq:T2ansatz}
 \begin{split}
 T^{\mu\nu} &= a_1 \eta^{\mu\nu} \partial_\rho\phi\partial^\rho\phi
 +a_2 \partial^\mu\phi \partial^\nu\phi
 +a_3 \eta^{\mu\nu} \phi \partial^2\phi
 +a_4 \phi \partial^\mu\partial^\nu \phi \,,
 \end{split}
\end{equation}
where the constant coefficients $a_i$ must be determined.
This tensor is already symmetric, so the requirement $T^{[\mu\nu]}=0$ does not fix any parameter (this requirement becomes nontrivial for $n\geq 2$). The ansatz, obviously, does not satisfy the remaining properties in \eqref{eq:main-conditions}. We find
\begin{equation}
 \begin{split}
 \partial_\mu T^{\mu\nu}|_{\rm on-shell} &= (2 a_1+a_2+a_4) \partial^\mu \phi \partial_\mu\partial^\nu \phi \,,
 \end{split}
\end{equation}
which, after requiring $\partial_\mu T^{\mu\nu}=0$, reduces the number of total parameters from four to three. We can solve the constraint in any of the three coefficients that appear above, e.g.~$a_4=-(2a_1+a_2)$.
Using the solution to the previous equation, we inspect the trace
\begin{equation}
 \begin{split}
 \delta_{\mu\nu} T^{\mu\nu}|_{\rm on-shell} &=
(a_2+d a_1) \partial_\mu \phi\, \partial^\mu \phi \,,
 \end{split}
 \end{equation}
which also reduces the number of parameters by one when set to zero, we take $a_2=-da_1$.
The third condition is the requirement that $T^{\mu\nu}$ is a primary, which gives in general
\begin{equation}
 \begin{split}
 b^\rho \hat{K}_\rho T^{\mu\nu}(0)|_{\rm on-shell} &=
 \big((d-2)a_1-a_4\big) b^\rho \eta^{\mu\nu} \phi \partial_\rho\phi
 + \big((d-2)a_2+d a_4\big) b^{(\mu} \phi \partial^{\nu)}\phi \,,
 \end{split}
\end{equation}
where the right hand side is evaluated at $x^\mu=0$ and is already zero when inserting the above formulas for $a_4$ and $a_2$.
In the case $n=1$, the condition that the operator is primary is not independent from the previous two: it can be used to determine two parameters, but it is easy to check explicitly
that the resulting equations are the same as those already provided by the previous two conditions,
so the on-shell tensor that is both conserved and traceless is automatically a primary in the case $n=1$.

Solving $a_2=-d a_1$ and $a_4=(d-2)a_1$, either using the first two conditions
or the requirement that the operator is a primary, we are left with a tensor that depends only on two parameters as expected from the considerations above
\begin{equation}\label{eq:T2ansatz-2}
 \begin{split}
 T^{\mu\nu} &= a_1 \big[
 (d-2)\phi \partial^\mu\partial^\nu \phi
 - d \partial^\mu\phi \partial^\nu\phi
 +\eta^{\mu\nu} \partial_\rho\phi\partial^\rho\phi
 \big]
 + a_3 \eta^{\mu\nu} \phi \partial^2\phi\,.
 \end{split}
\end{equation}
Obviously the parameter $a_3$ decouples on-shell because it multiplies a term proportional to the equations of motion
\begin{equation}\label{eq:T2ansatz-3}
 \begin{split}
 T^{\mu\nu}|_{\rm on-shell} &= a_1 \big[
 (d-2)\phi \partial^\mu\partial^\nu \phi
 - d \partial^\mu\phi \partial^\nu\phi
 +\eta^{\mu\nu} \partial_\rho\phi\partial^\rho\phi
 \big]\,,
 \end{split}
\end{equation}
so on-shell we are left with a tensor that is determined completely up to its normalization (e.g.~a single parameter, which is $a_1$ in this case.
The leftover parameter represents the normalization of the EMT operator of
the CFT.
For the EMT of a CFT the normalization cannot be chosen at will, as is
typically the case for other primary operators, due to a
Ward identity that fixes the coefficient of three-point
functions of operators with the EMT~\cite{Osborn:1993cr}.
The coefficient of the two-point function of
the EMT is commonly referred to as the central charge, $C_T$. For the
theories that we are considering a general expression for $C_T$ can be found
in~\cite{Osborn:2016bev, Guerrieri:2016whh}.

In order to completely fix any redundancy of $T_{\mu\nu}$, even off-shell,
we resort to the extra conditions \eqref{eq:extra-conditions}.
These, to some extent, ensure that the $T_{\mu\nu}$ comes from a variational principle of an action in curved space.
The off-shell divergence
\begin{equation}
 \partial_\mu T^{\mu\nu} =
 (a_3-da_1) \partial^\nu\phi \partial^2 \phi
 +(a_3+(d-2)a_1) \phi\partial^\nu \partial^2 \phi\,,
\end{equation}
should be proportional only to the equations of motion and not to their first derivative, implying $a_3=-(d-2)a_1$, which fixes the next to last parameter.
Finally, the off-shell trace
\begin{equation}
 \delta_{\mu\nu} T^{\mu\nu} = -a_1(d-1)(d-2) \phi \partial^2 \phi \,,
 \label{eq:off-shell-trace}
\end{equation}
is required to be proportional to $\delta_2=\frac{d-2}{2}$ as in \eqref{eq:extra-conditions}, that happens
only if
$a_1(d-1)(d-2) = - \frac{d-2}{2} $, and thus $a_1=- \frac{1}{2(d-1)}$.

Evaluating the tensor given in Eq.~\eqref{eq:T2ansatz-3}, that was completely fixed on-shell, using the extra conditions, we find
\begin{equation}\label{eq:T2-no-structure}
 \begin{split}
  T^{\mu\nu} &= \frac{1}{2}\frac{d}{d-1}\partial^\mu \phi \partial^\nu \phi
  -\frac{1}{2}\frac{d-2}{d-1}\phi\partial^\mu\partial^\nu\phi
  -\frac{1}{2}\frac{1}{d-1}\eta^{\mu\nu} \partial^\rho \phi\partial_\rho \phi
  +\frac{1}{2}\frac{d-2}{d-1}\eta^{\mu\nu}\phi\partial^2\phi\,.
 \end{split}
\end{equation}
With some algebra and using the equations of motion, we can rewrite it as
\begin{equation}\label{eq:T2-structure}
 \begin{split}
  T^{\mu\nu} &= \partial^\mu \phi \partial^\nu \phi
  -\frac{1}{2} \eta^{\mu\nu} \partial_\rho \phi \partial^\rho \phi
  -\frac{d-2}{4(d-1)} (\partial^\mu\partial^\nu -\eta^{\mu\nu}\partial^2 ) \phi^2\,,
 \end{split}
\end{equation}
which should have a more familiar form, as it can be checked to coincide with the tensor that comes from varying
the quadratic action \eqref{eq:general-action-sc} and operator \eqref{eq:Delta2}
in flat space.
The normalization as fixed by \eqref{eq:off-shell-trace} due
to \eqref{eq:extra-conditions} is
the one that gives the correct central charge in the two-point function of
the EMT.

Using our previous definition of the differential operator ${\cal P}$ in \eqref{eq:P-def}, we notice
\begin{equation}\label{eq:T2-more-structure}
 \begin{split}
  T^{\mu\nu} = \partial^\mu \phi \partial^\nu \phi
  -\frac{1}{2} \eta^{\mu\nu} \partial_\rho \phi \partial^\rho \phi
  -{\cal P}^{\mu\nu} {\cal O}\,,
  \qquad {\cal O}=\frac{d-2}{4}\phi^2\,,
 \end{split}
\end{equation}
that highlights the structure \eqref{eq:T-general-structure} of the EMT.
In the variational derivation,
the differential operator ${\cal P}$ comes about as the functional derivative
of the $Q$-curvature in the Yamabe operator
after integrating by parts in the flat space limit.
As expected for $n=1$ there are no tensor structures with two indices as there are no possible operators that can play that role.

\subsection{Results for \texorpdfstring{$1\leq n \leq 8$}{1<=n<=8}}

The procedure outlined for $T^{\mu\nu}$ in the case $n=1$ (two derivatives)
works equally well for determining $T^{\mu\nu}$ up to $n=8$ (sixteen derivatives).
The explicit form of the tensors $T^{\mu\nu}$ can be found in an ancillary Mathematica file with the submission of this paper.
The file uses heavily the Mathematica packages \texttt{xAct} \cite{xact,Brizuela:2008ra} and \texttt{xTras} \cite{Nutma:2013zea} and runs in a reasonable time on an average desktop computer.
We display the full form EMT for the case $n=5$ with ten derivatives in
Appendix \ref{appendix:d10}. The explicit form is given for two main
reasons: it is the first EMT that is completely new (the case $n=4$
on-shell has already appeared in \cite{Guerrieri:2016whh}), and it is
useful for a comparison with the results of Sec.~\ref{sect:d10-emt} in
which the same EMT is shown, but in a way that underlines some of its
structures. For higher values of $n$ an increasing importance goes to the
condition that $T^{\mu\nu}$ is a primary operator, because it fixes several
otherwise free parameters in the general ansatz. The results up to $n=8$
are summarized in Table~\ref{tableI}.
\begin{table}[H]
\begin{center}
\begin{tabular}{ | c || c | c | c | c | c | c | }
\hline
derivatives & tot.\ pars.\ & $T^{[\mu\nu]}$ & $\partial_\mu T^{\mu\nu}$ & $T^\mu{}_\mu$ & $\hat{K}_\rho T^{\mu\nu}$ & extra \\
\hline
\hline
$2$ & $4$ & $4$ & $3$ & $2$ & $2$ & $0$ \\
\hline
$4$ & $10$ & $9$ & $5$ & $2$ & $2$ & $0$ \\
\hline
$6$ & $18$ & $16$ & $8$ & $3$ & $2$ & $0$ \\
\hline
$8$ & $29$ & $25$ & $12$ & $4$ & $2$ & $0$ \\
\hline
$10$ & $42$ & $36$ & $17$ & $6$ & $2$ & $0$ \\
\hline
$12$ & $58$ & $49$ & $23$ & $8$ & $2$ & $0$ \\
\hline
$14$ & $76$ & $64$ & $30$ & $11$ & $2$ & $0$ \\
\hline
$16$ & $97$ & $81$ & $38$ & $14$ & $2$ & $0$ \\
\hline
\end{tabular}
\end{center}
\caption{In each row the numbers indicate the leftover parameters after
successive applications of the on- and off-shell conditions on the ansatz
of $T^{\mu\nu}$.  Notice that the two extra conditions on the off-shell
divergence and trace always fix the two parameters left over by the
application of the others.  These two parameters are always associated with
the coefficient of the term $\eta^{\mu\nu}\phi (\partial^2)^n\phi$, which
is not fixed on-shell, and the overall normalization of the other terms
(likewise the parameter $a_1$ in \eqref{eq:T2ansatz-3} for the $n=1$
example).\label{tableI}
}
\end{table}

An algorithm, at least in spirit similar to ours, has been applied before
to completely fix the on-shell EMT up to $n=4$ (eight derivatives) and was
summarized in a similar table in \cite{Guerrieri:2016whh}. The numbers of
our table differ by one when compared with \cite{Guerrieri:2016whh},
because our ansatze contain the aforementioned term $\eta^{\mu\nu}\phi
(\partial^2)^n \phi$, which is always zero on-shell, but appears in the
variational tensor off-shell.  Our algorithm seems to be more efficient
because we have required that the EMT is a primary operator directly,
instead of requiring that it behaves as the CFT's $T^{\mu\nu}$ inside three
point functions, which is computationally more intensive because it
necessitates several Wick contractions. The strength of the present
approach thus lies in the algorithm developed in \cite{Osborn:2016bev} and
summarized in Appendix \ref{appendix:K}

In order to gauge the relative strength of each constraint, we list in
Table \ref{tableII} the number of equations that arise by enforcing each
property. The total number of constraining equations is bigger than the sum
of the free parameters of each ansatz, because several equations are
linearly dependent (for example, enforcing that the EMT is a primary
operator is equivalent to requiring that it is traceless and conserved in
the example $n=1$).  From the table it is evident that the primary
constraint is the relatively most important one, because it always fixes
all but two parameters. These two parameters are, in fact, the ones that
need to be fixed with the extra conditions, and contain the normalization
of the EMT. They can be seen by subtracting the extra consitions off-shell
with the corresponding on-shell conditions, which give one condition each
from divergence and trace, respectively. These are labelled as ``extra
div'' and ``extra tr'' in Table \ref{tableII}. We comment on the
possibility of generalizing the results on this section to arbitrary $n$ in
Sec.~\ref{sect:generaln}.

One unfortunate property of brute forcing the ansatz, as we have done in this section, is that the final result does not naturally display the ``structure'' that the discussion of the previous section has forecasted in \eqref{eq:T-general-structure}.
In part, this happens because we are not developing the construction of $T^{\mu\nu}$
starting from a non-improved and non-primary tensor $T_c^{\mu\nu}$.
The difference between a structured and non-structured result is evident when comparing \eqref{eq:T2-structure} with \eqref{eq:T2-more-structure}, where in the former we have rewritten the latter such that the differential operator ${\cal P}^{\mu\nu}$ is manifest, which allows to read ${\cal O}=\frac{d-2}{4}\phi^2$
using \eqref{eq:T-general-structure}. The task of evidencing the structure beneath $T^{\mu\nu}$ is rather nontrivial for higher values of $n$ and has been discussed in \cite{Osborn:2016bev} up to $n=3$.
To amend the lack of structure in the above results, in the next section we show another algorithm that uses a special choice of $T_c^{\mu\nu}$ as starting point of the improvement.

\begin{table}[H]
\begin{center}
\begin{tabular}{ | c || c | c | c | c | c | c | c | }
\hline
derivatives & tot.\ pars.\ & $T^{[\mu\nu]}$ & $\partial_\mu T^{\mu\nu}$ & $T^\mu{}_\mu$ & $\hat{K}_\rho T^{\mu\nu}$ & extra div & extra tr\\
\hline
\hline
$2$ & $4$ & $0$ & $1$ & $1$ & $2$ & $2$ & $2$ \\
\hline
$4$ & $10$ & $1$ & $4$ & $3$ & $8$ & $5$ & $4$ \\
\hline
$6$ & $18$ & $2$ & $8$ & $5$ & $16$ & $9$ & $6$ \\
\hline
$8$ & $29$ & $4$ & $13$ & $8$ & $27$ & $14$ & $9$ \\
\hline
$10$ & $42$ & $6$ & $19$ & $11$ & $40$ & $20$ & $12$ \\
\hline
$12$ & $58$ & $9$ & $26$ & $15$ & $56$ & $27$ & $16$ \\
\hline
$14$ & $76$ & $12$ & $34$ & $19$ & $74$ & $35$ & $20$ \\
\hline
$16$ & $97$ & $16$ & $43$ & $24$ & $95$ & $44$ & $25$ \\
\hline
\end{tabular}
\end{center}
\caption{In each row we list the number of independent
constraints on the parameters of $T^{\mu\nu}$ as a result of the
corresponding condition.\label{tableII}}
\end{table}

\section{Stress energy from improvement of minimally coupled actions}
\label{sect:improvement-minimal}

As discussed in Sec.~\ref{sect:weyl}, in order to write the EMT with a manifest structure as displayed in \eqref{eq:T-general-structure}, it is convenient to have an initial $T_c^{\mu\nu}$ as starting point of the improvement procedure.
The starting point could be the canonical EMT obtained from the Lagrange equations and the requirement that the Lagrangian is independent from the coordinates. However, it is more convenient to start from a nonimproved EMT which is already symmetric, which requires only an improvement of the form of the tensor $Z^{\mu\nu}$ and we pursue this route in this section.

To begin with, we perform a minimal generalization to curved space of the higher-derivative actions in flat space with equations of motion
$(-\partial^2)^n\phi=0$.
For odd powers $n$ we take the Lagrangians to be in the form
${\cal L}=-(\partial_\mu \Box^{\frac{n-1}{2}}\phi)^2$,
while for even powers $n$ we take instead
${\cal L}=-(\Box^{\frac{n}{2}}\phi)^2$.
These Lagrangians can be minimally extended to curved space by replacing $\partial_\mu \to \nabla_\mu$, $\Box \to g^{\mu\nu}\nabla_\mu\nabla_\nu$, and multiplying ${\cal L}$
by the covariant volume element, $1\to \sqrt{-g}$. We denote by $S_{\rm m}[\phi]$ the minimal extension to curved space of the action with $n$ derivatives.

Using the minimal extension, we define the stress energy tensor
$T_{c}^{\mu\nu}
= \left.\frac{2}{\sqrt{-g}}\frac{\delta S_{\rm m}}{\delta g_{\mu\nu}}\right|_{\rm flat}$, for which we explicitly evaluate the result in flat space.
It is straightforward to show that $T_{c}^{\mu\nu}$
is symmetric and conserved $\partial_\mu T_{c}^{\mu\nu}=0 $ using the
equations of motion $(-\partial^2)^n \phi=0$ and the invariance of $S_m$
under diffeomorphisms. In general, the tensor is not traceless nor is it a
primary, so we have our natural starting point as in the second part of
Sec.~\ref{sect:weyl}.

To obtain the improved traceless tensor, it is sufficient to write its trace as
$\delta_{\mu\nu}T_{c}^{\mu\nu}=\partial_\mu\partial_\nu Z^{\mu\nu}$,
for some symmetric tensor $Z^{\mu\nu}$. In this case, we have that the new tensor
$T_i^{\mu\nu}
=T_{c}^{\mu\nu}+{\cal D}^{\mu\nu}{}_{\alpha\beta} Z^{\alpha\beta}$
is symmetric, conserved and traceless, so it can be the stress energy tensor of a CFT. It might not be a primary at this stage.

Given the most general parametrization to $Z^{\mu\nu}$, we see that
multiple parametric choices are possible, especially for higher values of $n$.
This freedom is divided in two main groups: on the one hand, we have the nontrivial kernel of ${\cal D}^{\mu\nu}{}_{\alpha\beta}$,
which can be used to simplify
the final form of the operators involved in writing down $Z^{\mu\nu}$.
On the other hand, we have an intrinsic
leftover parametric dependence, because we have not yet imposed that the EMT is a primary operator. The first freedom does not appear in the final form of $T_{\mu\nu}$, while the latter obviously does (for $n\geq 3$).

When a leftover dependence shows, we then recast the tensor part writing it in the form ${\cal D}_B^{\mu\nu}{}_{\alpha\beta} \tilde{S}^{\alpha\beta}$, for an appropriate $\tilde{S}^{\alpha\beta}$, which can also come from a general ansatz. The form of the operator $\tilde{S}^{\alpha\beta}$ can similarly be simplified using the null directions of
${\cal D}_B^{\mu\nu}{}_{\alpha\beta}$, which for arbitrary $n$ are present. The coefficient of this structure is then fixed by the requirement that the improved tensor $T_i^{\mu\nu}$ is a also primary.

Finally, after having removed from $T_i^{\mu\nu}$ the part proportional to ${\cal D}_B^{\mu\nu}{}_{\alpha\beta}$, we ensure that the rest is written as
$T^{\mu\nu} = T_{c}^{\mu\nu} + {\cal D}^{\mu\nu}{}_{\rho\theta}(
{\cal O}\eta^{\rho\theta}+{\cal S}^{\rho\theta})$.
As said, we use the null directions of
${\cal D}^{\mu\nu}{}_{\rho\theta}$ to simplify as much as possible the tensor
operator ${\cal S}^{\mu\nu}$. Combining all the steps together
and using \eqref{eq:P-algebra}, we can write
the primary improved EMT as
\begin{equation} \label{eq:general-emt-structure}
 \begin{split}
 T^{\mu\nu} &= T_{c}^{\mu\nu}
 - {\cal P}^{\mu\nu} {\cal O}_{2n}
 + {\cal D}^{\mu\nu}{}_{\rho\theta}{\cal S}_{2n}^{\rho\theta}
 + {\cal D}_B^{\mu\nu}{}_{\rho\theta}\tilde{\cal S}_{2n}^{\rho\theta}\,,
 \end{split}
\end{equation}
which has thus the same form as conjectured in \eqref{eq:T-general-structure}.

\subsection{Two derivative EMT for \texorpdfstring{$n=1$}{n=1}}

To illustrate the procedure, consider once again the case $n=1$. A simple computation reveals the variational tensor on the minimal Lagrangian ${\cal L}=-\frac{1}{2}\sqrt{-g}g^{\mu\nu}\partial_\mu\phi\partial_\nu\phi$ to be
\begin{equation}
 \begin{split}
 T_{c}^{\mu\nu} = \partial^\mu\phi\partial^\nu\phi-\frac{1}{2} \eta^{\mu\nu} \partial_\rho\phi\partial^\rho\phi
 \,.
 \end{split}
\end{equation}
For $n=1$ the kernel of ${\cal D}$ is trivial and we do not have enough derivatives to even need ${\cal D}_B$.
The most general ansatz for $Z^{\mu\nu}$ must contain two derivatives less than the tensor, so it is trivial $Z^{\mu\nu}=  c_1 \eta^{\mu\nu} \phi^2$, i.e.\ a symmetric tensor depending quadratically on the field $\phi$ with no derivatives acting on it. From the on-shell requirement that
$\delta_{\mu\nu}T_{c}^{\mu\nu}=\partial_\mu\partial_\nu Z^{\mu\nu}$
we find $c_1=\frac{2-d}{4}$.
Otherwise we could use the off-shell trace to deform the condition above.
The tensor $Z^{\mu\nu}$ is proportional to the metric and $T^{\mu\nu}$ is already a primary. Consequently, the only improvement that contributes is the scalar ${\cal O}=\frac{d-2}{4}\phi^2$ and the final result agrees with \eqref{eq:T2-more-structure}. This is essentially the same strategy that was followed in the early works involving the computation of the improved EMT for scalar theories \cite{Brown:1977pq}.

\subsection{Four derivative EMT for \texorpdfstring{$n=2$}{n=2}}
\label{eq:4d-min-decomposition}

For $n=2$ we take the covariant minimal action ${\cal L}=-\frac{1}{2}\sqrt{-g}(\Box\phi)^2$.
The variational tensor is
\begin{equation}\label{eq:T4c}
 \begin{split}
 T_{c}^{\mu\nu} = -2 \partial^2\partial^{(\mu}\phi\partial^{\nu)}\phi
 +\eta^{\mu\nu} \partial^2\partial^\rho\phi \partial_\rho\phi
 +\frac{1}{2}\eta^{\mu\nu} (\partial^2\phi)^2
 \,.
 \end{split}
\end{equation}
This time, the general ansatz for $Z^{\mu\nu}$ is more complicated, since it belongs to a vector space constructed on all possible terms with two fields $\phi$ and two derivatives:
\begin{equation}
 \begin{split}
 Z^{\mu\nu} = c_1 \eta^{\mu\nu} (\partial\phi)^2 + c_2 \partial^\mu\phi\partial^\nu\phi + c_3 \eta^{\mu\nu} \phi\partial^2\phi +c_4  \phi \partial^\mu\partial^\nu\phi
 \,.
 \end{split}
\end{equation}
Enforcing the relation $\delta_{\mu\nu}T_{c}^{\mu\nu}=\partial_\mu\partial_\nu Z^{\mu\nu}$ on-shell, we find $c_2=1-c_1$, $c_3=\frac{d-2}{2}+c_1$ and $c_4=-1-c_1$,
by solving for all independent structures in terms of the first coefficient $c_1$.
The dependence on $c_1$ is expected because for $n=2$ we see that ${\cal D}$, with a domain in this vector space,
has a one dimensional kernel proportional to the combination
\begin{equation}
 \begin{split}
 \eta^{\mu\nu} (\partial\phi)^2 - \partial^\mu\phi\partial^\nu\phi + \eta^{\mu\nu} \phi\partial^2\phi - \phi \partial^\mu\partial^\nu\phi
 \,.
 \end{split}
\end{equation}
As a consequence the final EMT does not depend on $c_1$. It is also already a primary, so we do not need ${\cal D}_B$. We can use $c_1$ to simplify the final form of the operators, though.
As a general strategy, we will always try to simplify
the final form of the tensor ${\cal S}^{\mu\nu}$ as much as possible.

Using the solutions for $c_i$ in $Z^{\mu\nu}$, we have
\begin{equation}
 \begin{split}
Z^{\mu\nu} = (1-c_1)\partial^\mu\phi\partial^\nu\phi
-(1+c_1)\phi \partial^\mu\partial^\nu \phi
+ c_1 \eta^{\mu\nu}(\partial\phi)^2
+\frac{d-2+2c_1}{2} \eta^{\mu\nu}\phi \partial^2\phi
 \end{split}
\end{equation}
which directly implies
\begin{equation}
 \begin{split}
{\cal O} =
 - c_1 (\partial\phi)^2
 -\frac{d-2+2c_1}{2} \phi \partial^2\phi\,,
 \qquad
 {\cal S}^{\mu\nu} = (1-c_1)\partial^\mu\phi\partial^\nu\phi
-(1+c_1)\phi \partial^\mu\partial^\nu \phi\,.
 \end{split}
\end{equation}
We always choose to simplify the terms in the symmetric tensors that are proportional to a single instance of $\phi$ with no derivatives.
With our prescription, the best choice
for $c_1$ simplifies ${\cal S}^{\mu\nu}$ to the first term, so we take
$c_1=-1$.
The end result is
\begin{equation}\label{eq:T4-parts}
 \begin{split}
 {\cal O} = (\partial\phi)^2-\frac{d-4}{2}\phi\partial^2\phi\,,
 \qquad {\cal S}^{\mu\nu}= 2 \partial^\mu\phi \partial^\nu\phi \,.
 \end{split}
\end{equation}
The full form of the EMT comes from
combining \eqref{eq:T4c} and \eqref{eq:T4-parts} in
\eqref{eq:general-emt-structure} to get the final improved primary EMT.
This final form, that we do not display for brevity, is considerably more
complicated than its parts, which
underlines in practice the convenience of expressing the geometrical structures on the EMT.
Note that an obstruction to the existence of such EMT in $d=1$ and $d=2$ is given by the poles hidden in the ${\cal P}$ and ${\cal D}$ differential operators.

\subsection{Six derivative EMT for \texorpdfstring{$n=3$}{n=3}}
\label{eq:6d-min-decomposition}

We take the minimal Lagrangian ${\cal L}=-\frac{1}{2}\sqrt{-g}g^{\mu\nu}\partial_\mu\Box\phi\partial_\nu\Box\phi$, the application of the variational principle gives
\begin{equation}
 \begin{split}
 T_{c}^{\mu\nu} =
 2 \partial^2\partial^2\partial^{(\mu}\phi\partial^{\nu)}\phi
 +\partial^\mu \partial^2\phi\partial^\nu \partial^2\phi
 -\eta^{\mu\nu} \partial^2\phi\partial^2\partial^2\phi
 -\eta^{\mu\nu}\partial^2\partial^2\partial^\rho\phi\partial_\rho\phi
 -\frac{1}{2} \eta^{\mu\nu} (\partial_\rho \partial^2\phi)^2
 \,.
 \end{split}
\end{equation}
The differential operator ${\cal D}$, acting on a vector space costructed with two fields and four derivatives, has a three dimensional kernel
(which we do not display for brevity),
while ${\cal D}_B$, with a domain spanned by terms with two fields and two derivatives,  has a one dimensional kernel proportional to
$$
\phi \partial^\mu\partial^\nu \phi + \partial^\mu\phi \partial^\nu \phi\,.
$$
In general,
we do not consider the two additional trivial elements of the kernel
of ${\cal D}_B$ that are proportional to the metric, which would also belong to the kernel of ${\cal D}_B$.
The general ansatz for $Z^{\mu\nu}$ has $9$ parameters (so it belongs to a vector space of dimension nine)
\begin{equation}
 \begin{split}
 Z^{\mu\nu}_4 &= c_1 \phi \partial^\mu\partial^\nu \partial^2 \phi
 +c_2 \eta^{\mu\nu } \phi \partial^2 \partial^2 \phi
 +c_3 \eta^{\mu\nu } \partial_\rho \phi \partial^\rho \partial^2  \phi
 +c_4 \partial_\rho \phi \partial^\rho \partial^\mu\partial^\nu  \phi
 +c_5 \partial^{(\mu} \phi \partial^{\nu)} \partial^2  \phi
 \\&
 \quad+c_6 \partial^\mu\partial^\rho\phi \partial^\nu\partial_\rho \phi
 +c_7 \eta^{\mu\nu } (\partial^2\phi)^2
 +c_8 \eta^{\mu\nu } \partial^\rho\partial^\lambda\phi \partial_\rho\partial_\lambda \phi
 +c_9 \partial^\mu\partial^\nu\phi \partial^2 \phi
 \,.
 \end{split}
\end{equation}
Of these parameters, only $5$ survive after imposing the condition
$\delta_{\mu\nu}T_{c}^{\mu\nu}=\partial_\mu\partial_\nu Z^{\mu\nu}$
on-shell. One solution of the $5$ remaining parameters in terms of the other $4$
is
\begin{equation}
 \begin{split}
 & c_2 = 3 - c_1 - \frac{d}{2}\,, c_5 = -4 - c_3 - c_4\,, c_7 = c_1 + c_3 + \frac{c_4+c_6}{2}  - \frac{d+6}{4}\,,
 \\&
 c_8 = -\frac{c_4+c_6}{2}\,, c_9 = 4 - c_1 - c_3 - c_6\,.
 \end{split}
\end{equation}
Of the $5$ parameters left, $3$ parametrize the kernel of ${\cal D}$
that can be used to simplify the final form at will later.
If improved in this way, $T^{\mu\nu}$ still retains a manifest dependence
on $2$ parameters left, which are $c_4$ and $c_6$ using our convention.
However, direct inspection of the improved tensor reveals that it depends only on the linear combination $c_4-c_6$. The part that multiplies the combination $c_4-c_6$ can be shown to be proportional to ${\cal D}_B$ acting on some symmetric
operator that we must determine to obtain $\tilde{\cal S}^{\mu\nu}$.
The presence of ${\cal D}_B$ is the crucial difference that the cases $n\geq 3$
have from the simpler $n=2$ shown before.

Using also the redundancy in the null directions of ${\cal D}_B$,
we arrange the improved tensor as in \cite{Osborn:2016bev} making manifest the $\lambda$ dependence
\begin{equation}
 \begin{split}
 T^{\mu\nu}=T_{c}^{\mu\nu}+{\cal D}^{\mu\nu}{}_{\rho\theta}({\cal O} \eta^{\rho\theta}+{\cal S}^{\rho\theta})
 +\lambda {\cal D}_B{}^{\mu\nu}{}_{\rho\theta}
 (\partial^\rho\phi\partial^\theta\phi)
 \end{split}
\end{equation}
for $\lambda=\frac{c_6-c_4}{2}$,
in agreement with the freedom of changing any improved stress energy tensor by a ${\cal D}_B$ contribution.
In general, there are multiple ways of writing $\tilde{\cal S}^{\rho\theta}$ due to the redundancy given by the kernel of ${\cal D}_B$. The general form of $\tilde{\cal S}^{\rho\theta}$ has two parameters, but one parametrizes the kernel of ${\cal D}_B$, while the normalization is hidden in $\lambda$.
Requiring that $T^{\mu\nu}$ is a primary, we find $\lambda=-\frac{8}{d-4}$, and thus we read
$\tilde{\cal S}^{\rho\theta}=-\frac{8}{d-4}\partial^\rho\phi\partial^\theta\phi$.

Having fixed the ${\cal D}_B$ contribution, we can similarly extract the combination ${\cal O} \eta^{\rho\theta}+{\cal S}^{\rho\theta}$, which can be simplified using the kernel of ${\cal D}$.
Collecting everything together, we find
\begin{align}
&{\cal O} = \frac{22+d}{4}(\partial^2\phi)^2
+ 4 \partial_\mu \partial^2\phi \partial^\mu\phi
+ \frac{d-6}{2}\phi \partial^2\partial^2 \phi\,,
&&\nonumber \\
&{\cal S}^{\mu\nu} =
8 \partial^\mu\partial^\nu\phi\partial^\rho\partial_\rho\phi\,,
&&\tilde{\cal S}^{\mu\nu}=-\frac{8}{d-4}\partial^\mu\phi\partial^\nu\phi\,,
 \end{align}
which can be plugged in \eqref{eq:T-general-structure} to have the explicit form
of the improved primary EMT. This EMT does not exist in $d=1,2,4$, because of the new pole in $\tilde{\cal S}$.

\subsection{Eight derivative EMT for \texorpdfstring{$n=4$}{n=4}}

For $n=4$ we take the covariant minimal action
${\cal L}=-\frac{1}{2}\sqrt{-g}(\Box^2\phi)^2$.
The variational tensor in flat space is
\begin{equation}
 \begin{split}
 T_{c}^{\mu\nu} =&
 -2\partial^{(\mu} \phi \partial^{\nu)} \partial^2\partial^2\partial^2\phi
 -2\partial^{(\mu} \partial^2 \phi \partial^{\nu)} \partial^2\partial^2\partial^2\phi
 +\eta^{\mu\nu}\partial^2\phi\partial^2\partial^2\partial^2\phi
 \\&
 +\frac{1}{2}\eta^{\mu\nu}\partial^2\partial^2\phi\partial^2\partial^2\phi
 +\eta^{\mu\nu}\partial_\rho\partial^2\phi\partial^\rho\partial^2\partial^2\phi
 +\eta^{\mu\nu}\partial_\rho\phi\partial^\rho\partial^2\partial^2\partial^2\phi
 \,.
 \end{split}
\end{equation}
The differential operator ${\cal D}$ has a six dimensional kernel, while
${\cal D}_B$ has a three dimensional one. Following the same steps as in the previous case we find the tensors
\begin{equation}
 \begin{split}
  {\cal O} =&
  2\partial^\mu \phi \partial_\mu \partial^2\partial^2\partial^2\phi
  +8\partial^\mu \partial^\nu \phi \partial_\mu \partial_\nu \partial^2\phi
  + 4 \partial_\mu\partial_\nu\partial_\rho\phi \partial^\mu\partial^\nu\partial^\rho\phi
  \\&
  -\frac{d-8}{2}\phi \partial^2\partial^2\partial^2\phi
  -\frac{d+8}{2}\partial^2\phi \partial^2\partial^2\phi
  \,,
  \\
  {\cal S}^{\mu\nu} =&
  -4 \partial^2\partial^2\phi \partial^\mu\partial^\nu \phi
  +8 \partial^\mu\partial^\nu\partial^\rho\phi \partial_\rho \partial^2 \phi
  +8 \partial^\mu \partial^\rho \partial^\theta \phi \partial^\nu \partial_\rho \partial_\theta \phi
  \,,
  \\
  \tilde{\cal S}^{\mu\nu} =&
  -\frac{8d}{d-6}\partial^\mu\partial^\nu\phi \partial^2\phi
  -\frac{4d(d-10)}{(d-4)(d-6)}\partial^\mu\partial^\rho\phi\partial^\nu\partial_\rho\phi
  \,.
 \end{split}
\end{equation}
Regardless of the parametric freedom in choosing $\tilde{\cal S}^{\mu\nu}$,
we always have that there are at least two independent tensor structures,
so one more than the case $n=3$. This can be evinced also from the results
of the table of Sec.~\ref{sect:ansatz}, where we see that the constraint
that the EMT is a primary is increasingly potent for higher values of $n$.
The form above is a compact (related to Weyl geometry) off-shell extension of the EMT which was computed with different methods in~\cite{Guerrieri:2016whh}. Note the new pole present in $d=6$.

\subsection{Ten derivative EMT for \texorpdfstring{$n=5$}{n=5}}\label{sect:10d}
\label{sect:d10-emt}

We take the minimal Lagrangian ${\cal L}=-\frac{1}{2}\sqrt{-g}g^{\mu\nu}\partial_\mu\Box^2\phi\partial_\nu\Box^2\phi$.
The variational principle gives
\begin{equation}
 \begin{split}
 &T_{c}^{\mu\nu} =
 2\partial^{(\mu} \phi \partial^{\nu)} \partial^2\partial^2\partial^2\partial^2\phi
 +2\partial^{(\mu} \partial^2 \phi \partial^{\nu)} \partial^2\partial^2\partial^2\phi
 +\partial^{\mu} \partial^2 \partial^2 \phi \partial^{\nu} \partial^2\partial^2\phi
 -\eta^{\mu\nu}\partial^2\phi\partial^2\partial^2\partial^2\partial^2\phi
 \\&\quad
 -\eta^{\mu\nu}\partial^2\partial^2\phi\partial^2\partial^2\partial^2\phi
 -\frac{1}{2}\eta^{\mu\nu} \partial_\rho\partial^2\partial^2\phi\partial_\rho\partial^2\partial^2\phi
 -\eta^{\mu\nu} \partial_\rho\partial^2\phi\partial_\rho\partial^2\partial^2\partial^2\phi
 -\eta^{\mu\nu} \partial_\rho\phi\partial_\rho\partial^2\partial^2\partial^2\partial^2\phi
 \,.
 \end{split}
\end{equation}
The differential operator ${\cal D}$ has a ten dimensional kernel, while
${\cal D}_B$ has a six dimensional one. Using the algorithm, the operators are
\begin{equation}\label{eq:d10-tensors}
 \begin{split}
  {\cal O} &=
  \frac{3d+2}{2} \partial^\mu \partial^2\partial^2\partial^2 \partial_\mu\phi
  -\frac{3d+26}{2} \partial^\mu \partial^2\partial^2 \phi \partial_\mu\partial^2\phi
  +\frac{3(d+6)}{2} \partial^\mu \partial^\nu \partial^2 \partial^2\phi \partial_\mu\partial_\nu\phi
  \\&
  \quad
  -6 \partial^\mu \partial^\nu \partial^2 \phi \partial_\mu\partial_\nu \partial^2\phi
  +\frac{3d+50}{2} \partial^\mu \partial^\nu \partial^\rho \partial^2 \phi \partial_\mu\partial_\nu \partial_\rho\phi
  -\frac{d+10}{2}\partial^2\partial^2\phi\partial^2\partial^2\phi
  \\&
  \quad
  +\frac{d+10}{2}\partial^2\phi \partial^2\partial^2\partial^2\phi
  +\frac{d-10}{2}\phi \partial^2\partial^2\partial^2\phi\partial^2\phi
  +\frac{3d+50}{4} \partial^\mu \partial^\nu \partial^\rho \partial^\theta \phi \partial_\mu\partial_\nu \partial_\rho \partial_\theta\phi
  \,,
  \\
  {\cal S}^{\mu\nu} &=
  -\frac{3d-14}{2} \partial^\mu\partial^\nu \partial^\rho \partial^2\partial^2\phi\partial_\rho\phi
  +\frac{3d+50}{2} \partial^\mu\partial^\nu \partial^\rho \partial^\theta \partial^2\phi \partial_\rho\partial_\theta\phi
  -12 \partial^\mu \partial^\rho \partial^2\phi \partial^\nu\partial_\rho\partial^2\phi
  \\&
  \quad
  +\frac{3d+50}{2} \partial^\mu \partial^\rho \partial^\theta \partial^\sigma\phi \partial^\nu\partial_\rho\partial_\theta\partial_\sigma\phi
  -\frac{3d+10}{2} \partial^\mu\partial^\nu \partial^2\phi \partial^2\partial^2\phi
  \,,
  \\
  \tilde{\cal S}^{\mu\nu} &=
  16\frac{d-2}{d-8}\partial^2\partial^2\phi\partial^\mu\partial^\nu\phi
  +\frac{3d^4-4d^3-588d^2+4624d-3456}{4(d-4)(d-6)(d-8)}\partial^\mu\partial^\nu\partial^\rho\partial^\theta\phi\partial_\rho \partial_\theta\phi
  \\&
  \quad-\frac{3d^3-10d^2+72d+64}{4(d-4)(d-8)}\partial^\mu\partial^2\phi\partial^\nu\partial^2\phi
  -\frac{3d^3+8d^2-556d+864}{2(d-6)(d-8)}\partial^\mu\partial^\nu\partial^\rho\phi\partial_\rho\partial^2\phi
  \,.
 \end{split}
\end{equation}
Once again we find consistency with the numbers of the table of Sec.~\ref{sect:ansatz} because we expect at least four independent tensor strutures remaining in $\tilde{\cal S}^{\mu\nu}$. A new pole appears in the even dimension $d=8$. The full $T^{\mu\nu}$ tensor for the case $n=5$
is displayed in Appendix~\ref{appendix:d10}. The complexity of the full
form should be compared with the result given above.

\subsection*{Comments on general \texorpdfstring{$n\geq2$}{n>=2}}
\label{sect:generaln}

The variational EMT for a minimal extension to curved space of the free scalar theory,
using the Lagrangian ${\cal L}=-\frac{1}{2}\phi \Box^n \phi$ can be found in \cite{Gibbons:2019lmj}, in which a method that involves the auxiliary scalars $\chi_k=\Box^k\phi$ for $k<n$ was adopted.
The final form of the nonimproved EMT is thus known,
but, expectedly, rather complicated.
Consider also that a different placement of the derivatives in ${\cal L}$ results in a different form for the final $T_c^{\mu\nu}$, but the most important
points of our algorithm still work independently of its specific form
for general $n$.

From the explicit formula for $T_c^{\mu\nu}$ in \cite{Gibbons:2019lmj},
one could envision
a general procedure that turns the nonimproved it
into an improved EMT $T_i^{\mu\nu}$, using a general form for $Z^{\mu\nu}$.
What is less clear is how difficult would be to enforce that $T_i^{\mu\nu}$
is a primary operator and extract the operators ${\cal O}$, ${\cal S}^{\mu\nu}$ and $\tilde{\cal S}^{\mu\nu}$ as in \eqref{eq:T-general-structure}, but in a general $n$-dependent form.

A recursion relation for the coefficients of a special class of primary operators
of arbitrary spin, known as ``single-trace'' operators, is known
\cite{Penedones:2010ue,Fitzpatrick:2011dm,Bekaert:2015tva,Brust:2016gjy},
but it has been solved only partially. A closed form is in fact known
only for the case $n=1$ \cite{Mikhailov:2002bp,Braun:2003rp,Osborn:2016bev}.
The use of the recursion relation is equivalent to enforcing the primary
constraint on the EMT, as we have done for both algorithms.
Even though the general form for the EMT could, in principle, be obtained,
it would still be a difficult task to rewrite it in the form
\eqref{eq:T-general-structure}, which originates from considerations
based on Weyl invariance in curved space.

\subsection*{Comments on the ``special'' dimensions}
\label{sect:speciald}

From the point of view of flat space CFT in arbitrary dimension,
the theories we are considering have two-point functions
\begin{equation}\label{eq:2pt-phi}
 \langle \phi(x) \phi(0) \rangle \sim \frac{1}{x^{2 \Delta_\phi}} \quad
 {\rm for}\quad \Delta_\phi=\delta_{2n}=\frac{d-2n}{2}\,,
\end{equation}
as a consequence of conformal invariance.  The scaling dimension of the
basic field $\phi$ is $\Delta_\phi$, which becomes negative for $d<2n$,
implying a positive power of $x^2$ in the two-point functions.
Nevertheless, the theories still make formal sense as CFTs
\cite{Brust:2016gjy}. In this context, they are understood as a particular
case of generalized free theories and their physical content is a
collection of operators with correlators satisfying CFT properties.
Intriguingly, in the special even dimensions $d\leq2n$ these CFTs have a
finite spectrum of operators \cite{Brust:2016gjy}.

For even $d\leq 2n$, which \emph{include} also the dimensions that appear
as poles in the computation of the EMTs and highlight the obstructions to
Weyl invariance, we have that \eqref{eq:2pt-phi} cannot be interpreted as
the Green's function of $\Box^n \phi$, i.e.\ the two-point functions are
not solutions of an equation of the form $\Box^n\langle \phi(x) \phi(0)
\rangle \sim \delta^{(d)}(x)$, but rather $\Box^n\langle \phi(x) \phi(0)
\rangle$ does not have a delta function and is either proportional to a
constant or to a positive power of $x^2$. From our field-theoretical
perspective this means that, in even $d\leq2n$, to obtain the CFTs
discussed in \cite{Brust:2016gjy} we are not allowed to go on-shell with
the relation $\Box^n \phi=0$.  As a consequence, we could still enforce
that a symmetric tensor, say $T'^{\mu\nu}$, is a primary operator using the
algorithm of Appendix~\ref{appendix:K}.  The resulting operator will depend
on two parameters, corresponding to the two normalization parameters
discussed in Sec.~\ref{sect:ansatz}, which for $d>2n$ were determined from
the extra conditions coming from the variational principle of the Weyl
action.  However, for $d\leq 2n$ we are not allowed to go on-shell, so if
we try to enforce the normalization of the EMT for the tensor $T'^{\mu\nu}$
we deduce that $T'^{\mu\nu}=0$, implying that conformal invariance is
incompatible with Weyl invariance.

A modification of \eqref{eq:2pt-phi} which restores the Green's function
property is to add logarithms, as discussed in \cite{Brust:2016gjy,
Farnsworth:2017tbz}, but logarithms are notoriously difficult to reconcile
with conformal invariance~\cite{Ferrara:1972jwi}.  It would be interesting
to pursue the interpretation of these limits as log-CFTs
\cite{Gurarie:1993xq}.  We do not pursue this idea further.

\section{Unitarity of Two- and Three-Point Functions of the
EMT}\label{sect:unitarity}

The higher-derivative theories that we are considering are nonunitary, but it is
interesting to ask if correlation functions of the EMT are sensitive to
this. For this question we will focus on $d=4$ and $d=6$, where concrete
results obtained in the literature for correlation functions of the EMT can
be used.

First of all, we know from \cite{Osborn:2016bev} that for $n=2$ the
two-point function of the EMT has negative coefficient $C_T$ and is thus
nonunitary, while for $n=3$ $C_T$ is positive. These statements are
$d$-independent. Moving on to the three-point function of the EMT, in $d=4$ we
may ask if the Hofman--Maldacena bounds~\cite{Hofman:2008ar} are satisfied
by our theory,\footnote{We thank Hugh Osborn for suggesting this.} namely
if the three-point function coefficients $a$ and $c$ are such that
\eqn{\frac13\leq\frac{a}{c}\leq\frac{31}{18}\,.}[hmbounds]
One could compute $a$ and $c$ using our results for specific $n$ in $d=4$,
but fortunately for $d=4$ $a$ and $c$ have been computed more generally~\cite{Bugini:2018def}:
\eqn{a=\frac{n^3}{144}-\frac{n^5}{240}\,,\qquad
c=a+\frac{n}{180}\,.}[]
Substituting these into \hmbounds we find that the bounds are in fact
satisfied for
$n\geq2$ and, as expected, the lower bound is saturated for $n=1$. It is
also easy to check that for $n\geq2$ it is $c<0$, which is consistent with
the result $C_T<0$ mentioned above for $d=4, n=2$, since, in our
conventions (namely those of \cite{Osborn:2016bev} and
\cite{Bugini:2018def}), $C_T=40c/\pi^4$.

In $d=6$ there are three coefficients in the three-point function of the
EMT, commonly denoted by $c_1, c_2, c_3$. The coefficient $c_3$ also
appears in the two-point function of the EMT.  Here we may use results of
\cite{Bugini:2018def},
\eqna{a&=-\frac{1}{7!}\frac{3n^7-21n^5+28n^3}{144}\,,\\
c_1&=96a+\tfrac{1}{7!}\tfrac89 n(7n^2-10)\,,\quad
c_2=24a-\tfrac{1}{7!}\tfrac29 n(7n^2-22)\,,\quad
c_3=-8a-\tfrac{1}{7!}\tfrac19 n(14n^2-27)\,,}[]
as well as \cite{deBoer:2009pn} and \cite{Bastianelli:2000hi} to show that the
associated bounds, which take the form
\eqn{C_1\equiv1-\tfrac15 t_2-\tfrac{2}{35}t_4\ge0\,,\qquad
C_2\equiv C_1+\tfrac12 t_2\ge0\,,\qquad
C_3\equiv C_1+\tfrac45(t_2+t_4)\geq0\,,}[Cineqs]
where $t_2$ and $t_4$ are related to the coefficients $c_1$, $c_2$
and $c_3$ by
\eqn{t_2=\frac{15\lsp(23\lsp c_1-44\lsp c_2+144\lsp c_3)}{16\lsp c_3}\,,
\qquad
t_4=-\frac{105\lsp(c_1-2\lsp c_2+6\lsp c_3)}{2\lsp c_3}\,,}[tres]
are satisfied for $n=1,2$ and $n\geq3$. We may also verify that $c_3>0$
(and thus the two-point function of the EMT is unitary) for $n=1$ and
$n\geq3$, but not for $n=2$. This $n=2, d=6$ result is again consistent with
the results of \cite{Osborn:2016bev}, which computed a negative $C_T$ when
$n=2$ for any $d$. We note that in $d=6$ we have the relation
$C_T=189c_3/\pi^6$ in the conventions of \cite{Osborn:2016bev} and
\cite{Bugini:2018def}.

We stress that the Hofman--Maldacena
bounds are just some of the necessary conditions
that correlators of unitary theories must satisfy.
They are not sufficient conditions, in fact the models that we consider are nonunitary theories for $n>2$ and the fact that some of their correlators are within the bounds does not imply the contrary.
For example, it is straightforward to see that the simplest unitarity constraints on the scaling dimensions can be satisfied also in nonunitary QFTs or CFTs, as is often encountered in the perturbative analysis of the anomalous dimension, which is positive for some nonunitary theories~\cite{Codello:2019isr}. If the fact that some bounds are satisfied
has a deeper meaning, for example if it could be used to construct a notion of distance in the space of CFTs between unitary and nonunitary theories, it is not clear.

\section{Generalization to Dirac spinors}\label{sect:dirac}

In Sect.~\ref{sect:weyl} we have briefly discussed the family of Weyl covariant
operators $\Delta_{2n}$ that act on scalars of dimension $\delta_{2n}=\frac{d-2n}{2}$. There exists a similar family of operators acting on Dirac spinors of dimension $\delta^s_{2n-1}=\frac{d-2n+1}{2}$, which we want do describe now.

In order to correctly couple Dirac spinors to curved space we need to introduce the vielbeins $e^a{}_\mu$, which are the components of a basis of forms $e^a=e^a{}_\mu{} dx^\mu$ that locally trivializes the metric $e^a{}_\mu{} e^b{}_\nu{} \eta_{ab}=g_{\mu\nu}$. The arbitrariness of constructing a local frame
introduces local Lorentz invariance, that acts on the Latin indices as $v^a \to \Lambda^a{}_b v^b$ with $\Lambda$ an $x^\mu$ dependent matrix. The compatible connection $\nabla_\mu$ can be extended to the spin frame
requiring compatibility with the vielbeins,
$\nabla_\mu e^a{}_\nu{}=\partial_\mu e^a{}_\nu -\Gamma_\mu{}^\rho{}_{\nu} e^a{}_\rho
+ A_\mu{}^a{}_b e^b{}_\nu=0$, from which one can deduce the components $A_\mu{}^a{}_b$ of the spin connection. The components are antisymmetric, $A_{\mu (a b)}=0$,
because they are Lie-algebra valued for any value of the coordinate index.

The Clifford algebra is introduced as in flat space,
thanks to the local frame.\footnote{%
This is the scholastic way of doing it. The reader interested in
circumventing this construction may consult \cite{Gies:2013noa,Lippoldt:2015cea}.
}
Since we do not specify the dimensionality $d$ of the manifold, the algebra can include an arbitrary large number of antisymmetric tensors with many indices.
The basic elements are the gamma metrices and satisfy $\{\gamma^a,\gamma^b\}=2\eta^{ab}$. Elements can be extended to coordinate indices defining $\gamma^\mu = e^\mu{}_a \gamma^a$, where $e^\mu{}_a$ are the inverse of the vielbeins (that is, a local orthonormal frame of vectors $e_a=e^\mu{}_a\partial_\mu$). Obviously, $\gamma^\mu$ depend on the coordinates.
Spinors transform under a representation of the Lorentz group with generators
$\Sigma^{ab}=\frac{1}{4}[\gamma^a,\gamma^b]$, so we can further extend the covariant derivative to their bundle as $\nabla_\mu \psi = (\partial_\mu+\frac{1}{2}A_{\mu a b}\Sigma^{ab})\psi$, for which we use the same symbol as the original connection to avoid overburdening the notation, and with some work one can check that this extension is covariant under both local transformations. We also define $\slashed{\nabla}=\gamma^\mu \nabla_\mu$ the familiar Dirac operator.

Since $d$ is not specified, the Clifford algebra can be arbitrarily large
and, in general, we must assume that antisymmetric products of gamma matrices with
arbitrarily many indices are present (specifying $d$ the algebra terminates
with the analog of $\gamma_5$ in general $d$, which occurs with $d$ or $d-1$ indices
depending on $d$ being even or odd). In the following, we need explicitly only
$\gamma^{\mu\nu\rho}$, which is defined
\begin{equation}
 \begin{split}
 \gamma^{\mu\nu\rho} &=
 \gamma^{[\mu} \gamma^\nu \gamma^{\rho]}= \frac{1}{6!} \gamma^{\mu} \gamma^\nu \gamma^{\rho} + {\rm perms.}
 \end{split}
\end{equation}
and satisfies
\begin{equation}\label{eq:gamma-algebra}
 \begin{split}
 &\gamma^\mu\gamma^\nu\gamma^\rho =
 \gamma^{\mu\nu\rho} + \gamma^\mu g^{\nu\rho}
 - \gamma^\nu g^{\mu\rho} + \gamma^\rho g^{\mu\nu}\,,\\
 &\gamma^{\mu}\gamma^{\nu\rho} =
 \gamma^{\mu\nu\rho} +\gamma^\rho g^{\mu\nu}-\gamma^\nu g^{\mu\rho} \,, \qquad
 \gamma^{\mu\nu}\gamma^{\rho} =
 \gamma^{\mu\nu\rho} +\gamma^\mu g^{\nu\rho}-\gamma^\nu g^{\mu\rho}\,.
 \end{split}
\end{equation}
The same relations hold with Latin indices provided the metric $g_{\mu\nu}$
is replaced with $\eta_{ab}$ of the local frame.

Weyl transformations are extended to the above construction requiring that the vielbein and its inverse transform as $e^a{}_\mu \to e^\sigma e^a{}_\mu $ and $e^\mu{}_a \to e^{-\sigma }e^\mu{}_a$, respectively, in agreement with the transformation of the metric. From the compatibility of the vielbein, one can deduce the transformation of the components $A_{\mu a b}$
\begin{equation}
 \begin{split}
 A_{\mu a b} \to A_{\mu a b} + 2 e^\nu{}_{[a} \eta_{b]c} e^c{}_{\mu} \partial_\nu \sigma\,.
 \end{split}
\end{equation}
Then, using \eqref{eq:gamma-algebra},
it is relatively easy to prove that the Dirac operator is a Weyl covariant operator,
$\slashed{\nabla} \to e^{-\frac{d+1}{2} \sigma} \slashed{\nabla} e^{\frac{d-1}{2}\sigma} $.
Consequently, the Lagrangian $-\overline{\psi}\slashed{\nabla}\psi$ is Weyl covariant with weight $d$, given a Dirac field $\psi$ and its conjugate $\overline{\psi}$ that transform with weight $\frac{d-1}{2}$, i.e.~$\psi \to e^{-\frac{d-1}{2}\sigma} \psi$ and $\overline{\psi} \to e^{-\frac{d-1}{2}\sigma} \overline{\psi}$.

The Dirac operator is actually the first element of a family of
anti-self-adjoint Weyl covariant operators $D_{2n-1}$ that transform
\begin{equation}
 \begin{split}
 D_{2n-1} \to D_{2n-1}'= e^{-\frac{d+2n-1}{2}\sigma} D_{2n-1} e^{\frac{d-2n+1}{2} \sigma}\,,
 \end{split}
\end{equation}
which reduces to the transformation of $\slashed{\nabla}$ for $n=1$, hence $D_1=\slashed{\nabla}$. In general, $D_{2n-1}$ are the unique Weyl covariant and Clifford algebra valued differential operators that satisfy the above transformation property \cite{holland-sparling}. The operators are rank-$(2n-1)$ and can be written in the form $D_{2n-1}= \slashed{\nabla}^{2n-1}+ \cdots$, where in the dots are hidden lower derivative terms and curvatures. In analogy to \eqref{eq:delta-operators-split}, we can separate the derivative and non-derivative parts
\begin{equation}\label{eq:D-operators-split}
 \begin{split}
  D_{2n-1} &= D_{2n-1,1} + \frac{d-2n+1}{2} {\cal Q}_{2n-1}\,,
 \end{split}
\end{equation}
where $D_{2n-1,1}$ is defined as the part of $D_{2n-1}$ that is zero on covariantly constant Dirac spinors $\psi_0$. The analog of the $Q$-curvatures,
here denoted ${\cal Q}_{2n-1}$, are Clifford algebra valued matrices that depend on Riemaniann curvatures. As such, they share similar transformation properties, but cannot directly be linked to topological invariants for a given conformal class.

The first element of the family is $D_1 = \slashed{\nabla}$ (for $n=1$), for which ${\cal Q}_1=0$, as we have already seen above (see also \cite{Kubo:1977ye}). The curvature ${\cal Q}_1$ could be inferred to be zero using dimensional analysis, because it has mass dimension one and Riemannian curvatures always have dimension two.\footnote{A nonzero value would necessitate torsion or nonmetricity to provide a covariant object of dimension one.}
The second element is for $n=2$ and has the form
\begin{equation}\label{eq:D3-operator}
 \begin{split}
  D_{3} &=
  \slashed{\nabla} \nabla^2+ \frac{2}{d-2}\gamma^\mu R_{\mu\nu} \nabla^\nu
  -\frac{d^2-3d+6}{4(d-1)(d-2)} R \slashed{\nabla}
  -\frac{d-3}{4(d-1)} (\slashed{\nabla}R)
  \,,
 \end{split}
\end{equation}
from which we read ${\cal Q}_{3}=\frac{1}{2(d-1)}\slashed{\nabla}R$.
Notice that, as expected, $D_{3} = \slashed{\nabla}^3+\cdots$ with the dots hiding
curvatures, because
we can always use the general relation $\slashed{\nabla}^2= \nabla^2 -\frac{R}{4}$
on Dirac spinors.
The explicit
forms of these operators have been written down up to $n=3$, that is, up to $D_5$ \cite{Fischmann-2013}. The operator $D_3$ for the case $d=4$ has appeared
before in the physics literature \cite{deBerredo-Peixoto:2001uob}, where its conformal anomaly was also computed.

Using the family of higher-derivative Weyl covariant operators $D_{2n-1}$ we can construct a family of Weyl invariant Dirac actions
\begin{equation}\label{eq:general-action-dirac}
 \begin{split}
   S[\psi,\overline{\psi},e] & = -\int d^{\lsp d}x \, \underline{e}\, \overline{\psi} D_{2n-1} \psi \,,
 \end{split}
\end{equation}
where we introduced the determinant of the vielbeins, $\underline{e}=\sqrt{-g}$.
By construction, the action is invariant under diffeomorphisms, local Lorentz transformations and Weyl transformations
\begin{align}
  &\delta^E_\xi e^a{}_\mu
  = \xi^\nu \partial_\nu e^a{}_\mu + e^a{}_\nu \partial_\mu \xi^\nu
  \,,
  &&\delta^E_\xi\psi = \xi^\mu \partial_\mu \psi
  \,,
  \\&
  \delta^W_\sigma e^a{}_\mu  = \sigma e^a{}_\mu
  \,,
  &&\delta^W_\sigma\psi = - \delta^s_{2n-1} \sigma \psi
  \,,
  \\&
  \delta^L_{\omega} e^a{}_\mu = \omega^a{}_b e^b{}_\mu
  \,,
  &&\delta^L_{\omega} \psi = \frac{1}{2} \omega_{ab}\Sigma^{a b} \psi
  \,,
\end{align}
where $\omega_{ab}$ is local and antisymmetric.
The on-shell equations for \eqref{eq:general-action-dirac} are
${\rvec{0.9ex}{D}}_{2n-1}\psi=0$ and $\overline{\psi} {\lvec{0.9ex}{D}}_{2n-1}=0$, which
reduce to ${\rvec{1ex}{{\slashed{\partial}}}}{}^{\,2n-1}\psi=0$ and
$\overline{\psi} {\lvec{1ex}{{\slashed{\partial}}}}{}^{\,2n-1}=0$
in flat space.

Using the Weyl covariant action we can define a (generalized) variational EMT
as the variation with respect to the vielbein
\begin{equation}\label{eq:generalized-T-spinors}
 \begin{split}
 T^{\mu}{}_a = \frac{1}{\underline{e}} \frac{\delta S}{\delta e^a{}_\mu}
 \,,\qquad
  T^{\mu\nu} = T^{\mu}{}_a \eta^{ab}e^\nu{}_b\,,
 \end{split}
\end{equation}
where in the second formula we have converted a frame index to a coordinate one using the inverse vielbein. This tensor is non expected to be symmetric in general, because we are not varying with respect a symmetric tensor. However, using the total functional differential as in \eqref{eq:total-diff-sc}
and the local Lorentz invariance, we can prove that $T^{\mu\nu}$ is symmetric
on-shell. Similarly it is conserved and traceless
\begin{equation}
 \begin{split}
 \nabla_\mu T^{\mu\nu}|_{\rm on-shell} = 0
 \,,\qquad
 g_{\mu\nu} T^{\mu\nu}|_{\rm on-shell}= 0\,,
 \end{split}
\end{equation}
so in the flat space limit it becomes the EMT of a CFT for the same arguments of Sec.~\ref{sect:weyl}. Off-shell we have
\begin{equation}
 \begin{split}
   \nabla_\mu T^{\mu\nu} = \overline{\psi}(
   \lvec{0.9ex}{D}_{2n-1}\rvec{0.9ex}{\nabla}^\nu-
 \lvec{0.9ex}{\hspace{0.5pt}\nabla}^\nu \rvec{0.9ex}{D}_{2n-1})\psi
 \,,\qquad
 g_{\mu\nu} T^{\mu\nu} = -(d-2n+1)
 \overline{\psi}\lrvec{0.9ex}{D}_{2n-1}\psi\,,
 \end{split}
\end{equation}
and
$\lrvec{0.9ex}{D}_{2n-1}=\frac{1}{2}(\rvec{0.9ex}{D}_{2n-1}-\lvec{0.9ex}{D}_{2n-1})$. The right hand sides are directly proportional to the equations of motion and not to their first derivatives, similarly to the scalar case.

If we restrict to the first special case, that is, to the traditional Dirac spinor, the variational EMT is already well-known in the literature.
As in Sec.~\ref{sect:ansatz}, we concentrate our attention to flat space,
$g_{\mu\nu}=\eta_{\mu\nu}$ and
$e^a{}_\mu=\delta^a_\mu$ for the local Lorentz group.
Using the algorithm of Sec.~\ref{sect:ansatz}, we could start from a general ansatz in flat space
\begin{equation}\label{eq:D1-T-ansatz}
 \begin{split}
 T^{\mu\nu} &= a_1 \overline{\psi} \gamma^\mu \partial^\nu \psi
 +a_2 \overline{\psi} \gamma^\nu \partial^\mu \psi
 +a_3 \partial^\mu \overline{\psi} \gamma^\nu \psi
 +a_4 \partial^\nu \overline{\psi} \gamma^\mu \psi
 +a_5 \eta^{\mu\nu}\overline{\psi} \slashed{\partial}\psi
 +a_6 \eta^{\mu\nu}\overline{\psi} \lvec{1ex}{\,\slashed{\partial}}\psi
 \\&\quad
 +a_7 \overline{\psi}\gamma^{\mu\nu\rho}\partial_\rho\psi
 +a_8 \overline{\psi}\gamma^{\mu\nu\rho}\lvec{0.9ex}{\,\partial}_\rho\psi\,,
 \end{split}
\end{equation}
which includes the Clifford algebra element $\gamma^{\mu\nu\rho}$, but does not need to include $\gamma^{\mu\nu}$ by virtue of \eqref{eq:gamma-algebra}.
Further gamma matrices with more than three indices can also be omitted because at least two indices must be contracted, but, by definition, they are antisymmetric, so $g_{\mu\nu}\gamma^{\mu\nu \alpha\cdots}=0$. The application of the algorithm of Sec.~\ref{sect:ansatz} reveals
that the symmetric conserved traceless primary tensor is
\begin{equation}
 \begin{split}
 T^{\mu\nu} &=  \overline{\psi}
 (\gamma^\mu \lrvec{0.9ex}{\hspace{0.75pt}\partial}^{\,\nu}-\eta^{\mu\nu}
   \lrvec{1ex}{\hspace{0.5pt}\slashed{\partial}\,}
 )\psi
 +\frac{1}{4}
 \overline{\psi}\gamma^{\mu\nu\rho}(\rvec{0.9ex}{\partial}_\rho
 +\lvec{0.9ex}{\,\partial}_\rho)\psi\,,
 \end{split}
\end{equation}
where $\lrvec{0.9ex}{\hspace{0.75pt}\partial}{}^{\,\nu}
=\frac{1}{2}(\rvec{0.9ex}{\partial}{}^{\,\nu}-\lvec{0.9ex}{\,\partial}{}^{\,\nu})$.
The above expression is symmetric \emph{on-shell}, as dictated by the invariance under local Lorentz transformations.\footnote{%
To go on-shell, we must use the fact that
$\gamma^{\mu\nu\rho}\partial_\rho \psi
= - \gamma^\mu \partial^\nu \psi + \gamma^\nu \partial^\mu \psi + \eta^{\mu\nu} \slashed{\partial}\psi$. A similar relation holds for $\overline{\psi}$.
}
Notice that the action of $\hat{K}_\mu$ is slightly different than for scalar operators in that it involves the Lorentz rotations generated by $\Sigma^{\mu\nu}$ and the algebra of the gamma matrices must be used extensively.

The above expression of $T^{\mu\nu}$ is probably unfamiliar, but can be
cast in a more standard form using the decomposition of
$\gamma^{\mu\nu\rho}$ in terms of gamma matrices with fewer indices
given in \eqref{eq:gamma-algebra}. One finds
\begin{equation}\label{eq:D1-T-final}
 \begin{split}
 T^{\mu\nu} &=  \frac{1}{2}\overline{\psi}
 (\gamma^\mu \lrvec{0.9ex}{\hspace{0.75pt}\partial}^{\,\nu}+\gamma^\nu
 \lrvec{0.9ex}{\hspace{0.75pt}\partial}^{\,\mu})\psi
 -  \eta^{\mu\nu} \overline{\psi}\,
 \lrvec{1ex}{\hspace{0.5pt}\slashed{\partial}}\, \psi
 +\frac{1}{2}\overline{\psi}(
 \lvec{1ex}{\,\slashed{\partial}} \,\Sigma^{\mu\nu}
 +\Sigma^{\mu\nu}\,\rvec{1ex}{\hspace{0.5pt}\slashed{\partial}}\,
 )\psi\,.
 \end{split}
\end{equation}
From the latest form, which coincides with the one found using a generalized N\"other approach~\cite{OsbornLectures} for conformal field theories, it is easy to recognize the terms involving the generator
of the Lorentz group ($\gamma^{\mu\nu}=2 \Sigma^{\mu\nu}$). If the computation of $T^{\mu\nu}$ is approached from the variational perspective as in \eqref{eq:generalized-T-spinors}, the last terms come about because of the spin connection, which is thus a fundamental ingredient in the construction of
the improved EMT (this can be inferred from the fact that it has a nontrivial Weyl transformation, that can be deduced from the compatibility relation of the vielbein and the transformation of the Christoffel symbols).

Moving on to a more complicated example, let us consider briefly the Dirac spinors with scaling dimension $\frac{d-3}{2}$.
The variational EMT coming from the higher-derivative action of \eqref{eq:D3-operator} has been considered before in \cite{amslaurea19852}. The same EMT has appeared also in \cite{Anselmi:1999bu} for $d=4$, where it was obtained with a similar logic to our algorithm (requiring tracelessness and conservation for the equations of motion
$\slashed{\partial}\partial^2\psi=0$) and the question on whether it came from a variational principle was posed. We now know the answer to be affirmative.
An ansatz such as the one of \eqref{eq:D1-T-ansatz} for the EMT of the Dirac
action with three derivatives is considerably more complicated than \eqref{eq:D1-T-ansatz}. The reason is that, since we do not want to assume a special value of $d$, there can be gamma matrices with an arbitrary number of indices. However, these are contracted with three partial derivatives and two indices must be left uncontracted, so the ansatz can contain at most the gamma matrices with five indices. The action of $\hat{K}_\mu$ introduces a further generator, which is proportional to the gamma matrices with two indices, so, in total, the algorithm
forces us to consider antisymmetric gamma matrices with up to seven indices that come from joining those with two and with five. In practice, the complete computation would require us to work in dimension $d>8$ to have the most general results, even though the operator \eqref{eq:D3-operator} ``lives'' naturally in $d\geq 3$.

Off-shell,
the symmetric part of the variational tensor \eqref{eq:generalized-T-spinors}
of the action \eqref{eq:general-action-dirac} with rank-$3$ $D_3$ operator \eqref{eq:D3-operator}
is
\begin{align}\label{eq:D3-T-final}
 &T^{(\mu\nu)} =
 \frac{1}{2}\overline{\psi} \gamma^{(\mu} \partial^{\nu)} \partial^2\psi
 -\frac{1}{2}\partial^{(\mu} \partial^2\overline{\psi} \gamma^{\nu)} \psi
 +\frac{1}{2} \partial_\rho \overline{\psi} \gamma^{\rho\theta(\mu} \partial_\theta \partial^{\nu)} \psi
 +\frac{1}{2}  \partial_\theta \partial^{(\mu}\overline{\psi} \gamma^{\nu)\rho\theta} \partial_\rho \psi
 \nonumber\\&\quad
 -\frac{d-3}{2(d-1)} \eta^{\mu\nu} \bigl(\overline{\psi} \gamma^\rho\partial^2 \partial_\rho\psi
 -\partial_\rho\partial^2 \overline{\psi} \gamma^\rho\psi
 \bigr)
 -\frac{1}{d-1} \bigl(
 \overline{\psi} \gamma^\rho \partial_\rho \partial^\mu\partial^\nu \psi
 -\partial_\rho \partial^\mu\partial^\nu\overline{\psi} \gamma^\rho \psi
 \bigr)
 \nonumber\\&\quad
 +\frac{d+2}{2(d-2)} \bigl(
 \partial^\rho \overline{\psi} \gamma^{(\mu} \partial^{\nu)} \partial_\rho \psi
 -\partial^\rho \partial^{(\mu} \overline{\psi} \gamma^{\nu)}  \partial_\rho \psi
 -\partial^{(\mu} \overline{\psi} \gamma^{\nu)} \partial^2\psi
 + \partial^2\overline{\psi} \gamma^{(\mu} \partial^{\nu)}\psi
 \bigr)
 \nonumber\\&\quad
 +\frac{d}{(d-1)(d-2)} \eta^{\mu\nu} \bigl(
 \partial_\rho \overline{\psi} \gamma^\rho \partial^2\psi
 - \partial^2\overline{\psi} \gamma^\rho \partial_\rho\psi
 \bigr)
 -\frac{2}{(d-1)(d-2)} \eta^{\mu\nu} \bigl(
 \partial_\theta \overline{\psi} \gamma^\rho \partial_\rho \partial^{\theta}\psi
 - \partial_\rho \partial^{\theta}\overline{\psi} \gamma^\rho \partial_\theta\psi
 \bigr)
 \nonumber\\&\quad
 -\frac{d^2-d+2}{2(d-1)(d-2)} \bigl(
 \partial_\rho \overline{\psi} \gamma^\rho \partial^\mu \partial^\nu \psi
 - \partial^\mu \partial^\nu\overline{\psi} \gamma^\rho  \partial_\rho\psi
 \bigr)
 \nonumber\\&\quad
 -\frac{d^2-3d-2}{2(d-1)(d-2)} \bigl(
 \partial^{(\mu} \overline{\psi} \gamma^{|\rho|} \partial_\rho \partial^{\nu)} \psi
 -\partial_\rho \partial^{(\mu} \overline{\psi} \gamma^{|\rho|}  \partial^{\nu)} \psi
 \bigr)\,,
\end{align}
evaluated for simplicity using a four-dimensional Clifford algebra.
We display only the symmetric part, because the antisymmetric one goes to zero on-shell. We also do not use the directional arrow notation because for more than one derivative it becomes rather cumbersome.
Using the on-shell relation $\slashed{\partial}\partial^2\psi=0$, it is possible
to further simplify the above structure.
The variational tensor on-shell was previously given in \cite{amslaurea19852}.

We can rewrite the above EMT following the decomposition \eqref{eq:general-emt-structure} using a ``minimal'' EMT as reference. For the computation of the minimal tensor we take
$S_{\rm m}=\frac{1}{2}\int \underline{e}(
\overline{\psi}\gamma^\mu \nabla_\mu \nabla^2 \psi - \nabla_\mu \nabla^2 \overline{\psi}\gamma^\mu  \psi)$,
which corresponds to an almost minimal extension to curved space of the theory with equations of motion $\slashed{\partial}\partial^2\psi=0$,
since we have made the kernel anti-self-adjoint.\footnote{%
Anti-self-adjointness of the kernel is necessary to have
an EMT that is anti-symmetric under the ``exchange'' of $\psi$ and $\overline{\psi}$. Otherwise, any conclusion that we are going to give
still applies even for other choices of $S_{\rm m}$.
}
The tensor $T_c^{\mu\nu}$ is practically defined as in Sec.~\ref{sect:improvement-minimal}, but we need to use the frame
$T_c^{\mu\nu}= \frac{1}{\underline{e}}\frac{\delta S_{\rm m}}{\delta e^a{}_\mu} e^{a\nu}$ as in \eqref{eq:generalized-T-spinors}. We have that $T_c^{\mu\nu}$
is conserved and symmetric on-shell, but it is not traceless because
the minimal extension to curved space is not Weyl invariant.
We can rewrite \eqref{eq:D3-T-final} as
\begin{equation}\label{eq:D3-T-decomposition}
 \begin{split}
 &T^{(\mu\nu)} = T_c^{(\mu\nu)} -{\cal P}^{\mu\nu}{\cal O} + {\cal D}^{\mu\nu}{}_{\rho\theta}{\cal S}^{\rho\theta}\,,
 \\
 &{\cal O} =
 -\frac{d-1}{2}\overline{\psi} \, \lrvec{1ex}{\hspace{0.5pt}\slashed{\partial}}\,  \psi \,,
 \qquad\qquad
 {\cal S}^{\mu\nu} =
 -2\overline{\psi} \gamma^{(\mu} \lrvec{1ex}{\hspace{0.5pt}\partial}^{\nu)} \psi\,,
 \end{split}
\end{equation}
from which we see that we do not need the ${\cal D}_B$ structure
(in analogy with the four derivative scalar case
shown in Sec.~\ref{eq:4d-min-decomposition}).
We are led to believe that the family of Weyl invariant Dirac actions
may be subject to similar obstructions to the scalar case.
Furthermore, we do not write down $T_c^{(\mu\nu)}$ for brevity: the tensor is quite
complicate because it contains all structures that already
appeared in \eqref{eq:D3-T-final}, but with $d$-independent coefficients;
its complete form can be found in the ancillary files to the submission.
We also conjecture that the ${\cal D}_B$ structure will appear starting
from the operator $D_5$ of the Dirac family (in analogy with the
six-derivative scalar case of Sec.~\ref{eq:6d-min-decomposition}). This is
supported by the explicit form of $D_5$ that has been computed
in~\cite[Theorem 5.49]{Fischmann2013} and contains a $1/(d-4)$ pole.

The computations of this section
have been checked also with the aid
of the Mathematica package \texttt{FieldsX} \cite{Frob:2020gdh},
which complements \texttt{xAct} with frame and spin structures
in arbitrary $d$. We have not checked that \eqref{eq:D3-T-final}
is the unique primary EMT with three derivatives, nor we have obtained it from
an ansatz procedure such as the one adopted for \eqref{eq:D1-T-final},
but we have checked that it is traceless, conserved and primary.

\section{Conclusion}\label{sect:conclusion}

We have approached the discussion of the energy momentum tensor of
conformal field theories from the point of view of curved space's Weyl
invariance. There is an intimate relation between Weyl and conformal
invariances: the former implies the latter, while the latter can be
extended to the former only if special geometrical obstructions are
circumvented.  These obstructions, that take the form of Weyl covariant
operators in even dimensions, appear in several contexts, including both
physics and differential geometry.

As toy models for our discussion, we have chosen two families of
higher-derivative free theories: a well-known one based on scalars, and a
less-known one based on Dirac spinors. In flat space both families
are nonunitary conformal field theories which exist in dimension $d>2$.
Despite their apparent simplicity, higher-derivative free theories are
quite nontrivial, a fact that can be deduced by directly inspecting their
Weyl invariant formulations and their energy momentum tensors.

The obstructions to Weyl invariance appear as poles in the dimension $d$ of
the corresponding energy momentum tensors. Using rather general
assumptions, we have argued a general form for the flat space limit of the
energy momentum tensor of higher-derivative free theories. From the results
of our analysis, we deduced that the only possible obstructions to the
construction of the conformal field theory's energy momentum tensor are
precisely the same ones that are obstructing the lift from conformal to Weyl
invariance.

Our discussion has been complemented with a variety of explicit
computations.  We have introduced first an algorithm that computes the
primary energy momentum tensor by ``brute-force'', and normalizes it using
relations that can be derived from the Weyl covariance analysis in curved space.
The results confirm the appearance of obstructions to Weyl invariance
for positive even
$d$.  We have also developed an algorithm that ``improves'' naive energy
momentum tensors of the higher-derivative free conformal field theories by
making them primary and traceless. In this second case, we have been able
to isolate not only the poles in $d$ of the obstructions, but also the
differential structures which we argue to be coming directly from the
obstructions themselves. As specific examples, we computed explicitly the
improved conformal primary energy momentum tensor for scalar theories with
a kinetic term having up to $16$ derivatives and reported compact
(geometrically inspired) expressions for the cases with up to $10$
derivatives.

A result of our considerations is a general finite ``decomposition'' of
the energy momentum tensor, which is based on the action of differential
operators on a scalar and two symmetric tensors.  We observed that the
decomposition is completely fixed if and only if the stress energy tensor
is a primary operator of the conformal field theory.

The requirement that an operator is a conformal primary is enforced through
the action of the generator of special conformal transformations in flat
space.  Instead, from the point of view of the Weyl covariant formalism,
the energy momentum tensor is the flat space limit of the functional
derivative of a Weyl invariant action with respect to the metric and there
is no obvious way to generalize the generator of special conformal
transformations to curved space. However, varying a Weyl invariant quantity
with respect to the metric gives a Weyl covariant tensor in curved space.
This implies that the corresponding flat-space version of this tensor will
be a conformal primary operator.

In this work our focus was on higher-derivative nonunitary free CFTs.
Naturally one may wonder how the picture presented could be extended when
interactions are present. There have recently been some perturbative
investigations (in the $\varepsilon$-expansion) devoted to the study of
several families of critical theories which are CFTs~\cite{Gliozzi:2016ysv,
Gracey:2017erc,Safari:2017irw, Safari:2017tgs, Safari:2021ocb}.  Therefore it would be
very interesting to push further the analysis of these conformal theories
and look at the relation to Weyl invariance in curved space, including also
the possible existence of obstructions.


\smallskip
\smallskip

\paragraph*{Acknowledgements.}
We thank H.\ Osborn for useful discussions and suggestions. OZ is grateful
to M.~Fr\"ob for helpful correspondence and suggestions on how to properly
set up computations with the package \texttt{FieldsX}. AS is funded by the
Royal Society under the grant ``Advancing the Conformal Bootstrap Program
in Three and Four Dimensions.''

\appendix

\section{Algorithm for \texorpdfstring{$\bm{\hat{K}_\mu}$}{K\_mu}} \label{appendix:K}

The algorithm to compute the action of the generators of the special conformal transformations on an arbitrary tensor constructed with derivatives acting on $\phi$ has been described in \cite{Osborn:2016bev} and we briefly summarize it here extending it to Dirac fermions.

Consider a local tensor $X_{\alpha_1 \cdots \alpha_m}$ in flat space that is constructed with partial derivatives, flat space's metrics and copies of the scalar field $\phi$ evaluated at the origin $x^\mu=0$. In general, every formula in this appendix should be regarded as evaluated at the origin. We first construct the operator $\hat{D}$ that counts the canonical mass dimension of $X_{\alpha_1 \cdots \alpha_m}$ as follows
\begin{equation}
 \begin{split}
 \hat{D} X_{\alpha_1 \cdots \alpha_m} = \Delta_X X_{\alpha_1 \cdots \alpha_m}
 \qquad {\rm for }~ X \sim \partial^{p_1} \phi^{p_2}\,,
 \end{split}
\end{equation}
where $\Delta_X=p_1 + \delta_{2n} p_2$ for $p_i$ are positive natural numbers and we recall that $\delta_{2n}$ is the canonical dimension of $\phi$.
Consequently, we have $\hat{D} \partial_\mu = \partial_\mu (\hat{D}+1)$ that allows for iterative applications.

Introducing a constant vector $b_\mu$, we have that the action of $\hat{K}_b\equiv b^\mu\hat{K}_\mu$ obeys the Leibnitz rule, thus we can concentrate our attention
on single monomials in which derivatives act on operators. The action on a general tensor can be deduced by commuting iteratively $\hat{K}_b$ with the partial derivatives
\begin{equation}\label{eq:K-action}
 \begin{split}
 [\hat{K}_b,\partial_\mu]X_{\alpha_1 \cdots \alpha_m} =
 \hat{D}X_{\alpha_1 \cdots \alpha_m}
 + \sum_{i=1}^m\{b_{\alpha_i} X_{\alpha_i\cdots \mu \cdots \alpha_m}
 -\delta_{\mu\alpha_i}b^\beta X_{\alpha_i\cdots \beta \cdots \alpha_m}
 \}\,.
 \end{split}
\end{equation}
The iteration of $\hat{K}_b$ lowers the number of derivatives by one, e.g.~on structures of the form $\partial^m \phi$ we have $\hat{K}_b\partial^m \phi \sim \partial^{m-1} \phi$. As a consequence, it acts on $\phi$ itself after $m$ applications or earlier. The operator $\phi$ is already a primary, so the algorithm always stops when $\hat{K}_b\phi=0$.

If we now introduce Dirac fields of dimension $\delta^s_{2n-1}$, it is
straightforward to generalize the action of $\hat{D}$,
e.g.~$ \hat{D}\psi=\delta^s_{2n-1}\psi$.
The operator $\hat{K}_b$ is now sensitive to the internal space of the spinors, however, because they transform nontrivially under the Lorentz group. The action on coordinate indices is unchanged when compared to \eqref{eq:K-action}, so, for illustrative purposes, we take a new operator $X$
that only has Dirac indices and conjugate companions. It is convenient to make the indices explict, $\psi \to \psi_A$ and $\overline{\psi}\to \overline{\psi}{}^A$. Let the operator $X_{B_1 \cdots B_q}^{A_1 \cdots A_p}$, we have
\begin{equation}\label{eq:K-action-dirac}
 \begin{split}
 [\hat{K}_b,\partial_\mu]X_{B_1 \cdots B_q}^{A_1 \cdots A_p} =
 \hat{D}X_{B_1 \cdots B_q}^{A_1 \cdots A_p}
 + b^\lambda \sum_{i=1}^q [\Sigma_{\lambda\mu}]_{B_i}{}^{C} X_{B_1 \cdots C \cdots B_q}^{A_1 \cdots A_p}
 - b^\lambda \sum_{i=1}^p X_{B_1 \cdots B_q}^{A_1 \cdots C \cdots A_p} [\Sigma_{\lambda\mu}]_C{}^{A_i} \,,
 \end{split}
\end{equation}
and $\Sigma_{\mu\nu}=\frac{1}{2}\gamma_{\mu\nu}=\frac{1}{4}[\gamma_\mu\gamma_\nu]$
are the generators of Lorentz transformations.
The generalization to operators $X$ with Greek indices includes the second term of \eqref{eq:K-action} and should be straightforward.
The terms involving $\gamma^{\mu\nu}$ are necessary to show that EMTs constructed with Dirac field, like for example \eqref{eq:D1-T-final}, are primary operators.

\section{EMT for ten derivatives} \label{appendix:d10}

We provide here the off-shell energy momentum tensor
for the higher-derivative free theory with ten derivatives
in its full ``unstructured'' form computed with the method of section \ref{sect:ansatz}.
It is possible to check that it coincides
with the results given in section \ref{sect:10d}
by using \eqref{eq:d10-tensors} in \eqref{eq:general-emt-structure}.
This is the first completely new tensor that we have computed, since the one with eight derivatives appeared in \cite{Guerrieri:2016whh} on-shell. The size of further EMTs for higher-derivatives
increases rapidly.
\begin{align}
  &T^{\mu\nu}
  =
  -\frac{d+8}{(d-1)} \partial^{(\mu} \partial^8 \phi \partial^{\nu)} \phi
  -\frac{(d+4)(d+8)}{(d-1)(d-8)} \partial^{(\mu} \partial^6 \phi \partial^{\nu)} \partial^2 \phi
  -\frac{d(d+4)(d+8)}{2(d-1)(d-4)(d-8)} \partial^{\mu} \partial^4 \phi \partial^{\nu} \partial^4 \phi
  \nonumber\\&
  -\frac{16(d+6)}{(d-1)(d-8)} \partial^{(\mu} \partial_\rho \partial^6\phi \partial^{\nu)}\partial^\rho\phi
  -\frac{192(d+4)}{(d-1)(d-6)(d-8)} \partial^{(\mu} \partial_\rho \partial_\sigma \partial^4\phi \partial^{\nu)}\partial^\rho \partial^\sigma\phi
  \nonumber\\&
  +\frac{24(d+2)(d^2+2d-32)}{(d-1)(d-4)(d-6)(d-8)} \partial^{(\mu} \partial_\rho\partial^4\phi \partial^{\nu)} \partial^\rho \partial^2\phi
  \nonumber\\&
  -\frac{1536(d+2)}{(d-1)(d-4)(d-6)(d-8)} \partial^{(\mu} \partial_\rho \partial_\sigma \partial_\theta \partial^2 \phi \partial^{\nu)}\partial^\rho \partial^\sigma \partial^\theta \phi
  \nonumber\\&
  -\frac{144d(d^2+2d-16)}{(d-1)(d-2)(d-4)(d-6)(d-8)} \partial^\mu \partial_\rho \partial_\sigma\partial^2\phi \partial^\nu \partial^\rho \partial^\sigma\partial^2\phi
  \nonumber\\&
  -\frac{3072d}{(d-1)(d-2)(d-4)(d-6)(d-8)}
  \partial^{\mu} \partial_\rho \partial_\sigma \partial_\theta \partial_\zeta \phi \partial^{\nu}\partial^\rho \partial^\sigma \partial^\theta \partial^\zeta \phi
  \nonumber\\&
  +\frac{(d+6)(d+8)}{2(d-1)(d-8)} \partial^\mu\partial^\nu\phi \partial^8\phi
  +\frac{8(d+4)(d+6)}{(d-1)(d-6)(d-8)} \partial^\mu\partial^\nu \partial_\rho\phi \partial^6\partial^\rho\phi
  \nonumber\\&
  +\frac{(d+2)(d+4)(d+8)}{2(d-1)(d-4)(d-8)}\partial^\mu\partial^\nu \partial^2\phi \partial^6\phi
  + \frac{96(d+2)(d+4)}{(d-1)(d-4)(d-6)(d-8)}\partial^\mu\partial^\nu \partial_\rho \partial_\theta \phi \partial^4\partial^\rho\partial^\theta\phi
  \nonumber\\&
  +\frac{768d(d+2)}{(d-1)(d-2)(d-4)(d-6)(d-8)} \partial^{\mu}\partial^\nu \partial^\rho \partial^\sigma \partial^\theta  \phi \partial_\rho \partial_\sigma \partial_\theta \partial^2 \phi
  \nonumber\\&
  +\frac{144(d^2+2d-16)}{(d-1)(d-4)(d-6)(d-8)}\partial^{\mu}\partial^\nu \partial^\rho \partial^\sigma \partial^2  \phi \partial_\rho \partial_\sigma  \partial^2 \phi
  \nonumber\\&
  +\frac{3072}{(d-1)(d-4)(d-6)(d-8)} \partial^\mu\partial^\nu\partial_\rho \partial_\theta\partial_\sigma \partial_\zeta\phi \partial^\rho \partial^\theta\partial^\sigma \partial^\zeta\phi
  \nonumber\\&
  +\frac{768}{(d-1)(d-6)(d-8)} \partial^\mu\partial^\nu \partial_\rho \partial_\sigma \partial_\theta \partial^2\phi \partial^\rho \partial^\sigma \partial^\theta \phi
  +\frac{96}{(d-1)(d-8)} \partial^\mu\partial^\nu \partial_\rho\partial_\sigma \partial^4 \phi \partial^\rho\partial^\sigma \phi
  \nonumber\\&
  +\frac{(d-2)(d+4)(d+8)}{2(d-1)(d-4)(d-8)} \partial^\mu\partial^\nu \partial^4\phi \partial^4\phi
  +\frac{d-10}{2(d-1)} \phi \partial^\mu\partial^\nu \partial^8\phi
  +\frac{(d-6)(d+8)}{2(d-1)(d-8)} \partial^\mu\partial^\nu \partial^6\phi \partial^2\phi
  \nonumber\\&
  +\frac{8}{d-1} \partial^\mu\partial^\nu \partial_\rho\partial^6\phi \partial^\rho\phi
  +\frac{12d(d+2)(d^2+2d-32)}{(d-1)(d-2)(d-4)(d-6)(d-8)} \partial^\mu\partial^\nu \partial_\rho \partial^2\phi \partial^\rho \partial^4\phi
  \nonumber\\&
  +\frac{12(d^2+2d-32)}{(d-1)(d-6)(d-8)} \partial^\mu\partial^\nu \partial_\rho \partial^4 \phi \partial^\rho \partial^2\phi
  -\frac{d-10}{2(d-1)} \eta^{\mu\nu} \phi \partial^{10}\phi
  -\frac{d+8}{(d-1)(d-8)} \eta^{\mu\nu} \partial^2\phi \partial^8\phi
  \nonumber\\&
  -\frac{(d+4)(d+8)}{(d-1)(d-4)(d-8)} \eta^{\mu\nu} \partial^4\phi \partial^6\phi
  +\frac{1}{d-1} \eta^{\mu\nu} \partial^\rho \partial^8\phi \partial^\rho\phi
  +\frac{d^2-14d-144}{(d-1)(d-6)(d-8)} \eta^{\mu\nu} \partial^\rho \partial^6\phi \partial^\rho \partial^2\phi
  \nonumber\\&
  +\frac{d^4-20d^3-148d^2+560d+1920}{2(d-1)(d-2)(d-4)(d-6)(d-8)}\eta^{\mu\nu} \partial^\rho \partial^4\phi \partial^\rho \partial^4\phi
  +\frac{16}{(d-1)(d-8)}\eta^{\mu\nu} \partial^\rho \partial^\sigma \partial^6\phi \partial^\rho \partial^\sigma \phi
  \nonumber\\&
  +\frac{24(d^2-6d-64)}{(d-1)(d-4)(d-6)(d-8)}\eta^{\mu\nu} \partial^\rho \partial^\sigma \partial^4\phi \partial^\rho \partial^\sigma \partial^2\phi
  \nonumber\\&
  +\frac{48(3d^2-10d-80)}{(d-1)(d-2)(d-4)(d-6)(d-8)}\eta^{\mu\nu} \partial^\rho \partial^\sigma \partial^\theta \partial^2\phi \partial_\rho \partial_\sigma \partial_\theta \partial^2\phi
  \nonumber\\&
  +\frac{192}{(d-1)(d-6)(d-8)}\eta^{\mu\nu} \partial^\rho \partial^\sigma \partial^\theta \partial^4\phi \partial_\rho \partial_\sigma \partial_\theta \phi
  \nonumber\\&
  +\frac{1536}{(d-1)(d-4)(d-6)(d-8)}\eta^{\mu\nu} \partial^\rho \partial^\sigma \partial^\theta \partial^\zeta\partial^2\phi \partial_\rho \partial_\sigma \partial_\theta \partial_\zeta \phi
  \nonumber\\&
  +\frac{3072}{(d-1)(d-2)(d-4)(d-6)(d-8)}
  \eta^{\mu\nu} \partial^\rho \partial^\sigma \partial^\theta \partial^\zeta \partial^\lambda \phi \partial_\rho \partial_\sigma \partial_\theta \partial_\zeta \partial_\lambda\phi
  \,.
\end{align}
We have compactly denoted powers of the Laplacian as $\partial^{2n}=
(\partial^2)^n$.  The above tensor illustrates how much more efficient is
the structured form \eqref{eq:d10-tensors} in comparison to the one
appearing here. Nevertheless, the above tensor still displays the poles for
all even dimensions in the interval $2\leq d\leq 8$ and for $d=1$, which
are related to obstructions when promoting Weyl invariance to full
conformal invariance of the EMT in lower dimensions.

\bibliographystyle{chetref}
\bibliography{biblio}

\begin{thebibliography}{10}
\ifx\href\asklfhas\newcommand{\href}[2]{#2}\fi
\ifx\arxivref\asklfhas\newcommand{\arxivref}[2]{\href{http://arxiv.org/abs/#1}{#2}}\fi
\ifx\doiref\asklfhas\newcommand{\doiref}[2]{\href{http://dx.doi.org/#1}{#2}}\fi
\parskip 0pt
\normalsize

\bibitem{Osborn:2016bev}
H.~Osborn \& A.~Stergiou,
\textit{``{C$_{T}$ for non-unitary CFTs in higher dimensions}''},
\doiref{10.1007/JHEP06(2016)079}{JHEP \textbf{1606}, 079
  (2016)\ignorespaces}\ignorespaces,
\normalsize{\texttt{\arxivref{1603.07307}{arXiv:1603.07307
  \![hep-th]}}}\ignorespaces.

\bibitem{Guerrieri:2016whh}
A.~Guerrieri, A.~C. Petkou \& C.~Wen,
\textit{``{The free $\sigma$CFTs}''},
\doiref{10.1007/JHEP09(2016)019}{JHEP \textbf{1609}, 019
  (2016)\ignorespaces}\ignorespaces,
\normalsize{\texttt{\arxivref{1604.07310}{arXiv:1604.07310
  \![hep-th]}}}\ignorespaces.

\bibitem{Brust:2016gjy}
C.~Brust \& K.~Hinterbichler,
\textit{``{Free \ensuremath{\square}$^{k}$ scalar conformal field theory}''},
\doiref{10.1007/JHEP02(2017)066}{JHEP \textbf{1702}, 066
  (2017)\ignorespaces}\ignorespaces,
\normalsize{\texttt{\arxivref{1607.07439}{arXiv:1607.07439
  \![hep-th]}}}\ignorespaces.

\bibitem{Nakayama:2016dby}
Y.~Nakayama,
\textit{``{Hidden global conformal symmetry without Virasoro extension in
  theory of elasticity}''},
\doiref{10.1016/j.aop.2016.06.010}{Annals~Phys. \textbf{372}, 392
  (2016)\ignorespaces}\ignorespaces,
\normalsize{\texttt{\arxivref{1604.00810}{arXiv:1604.00810
  \![hep-th]}}}\ignorespaces.

\bibitem{Graham-existence}
C.~R. Graham, R.~Jenne, L.~J. Mason \& G.~A.~J. Sparling,
\textit{``{Conformally Invariant Powers of the Laplacian, I: Existence}''},
\doiref{10.1112/jlms/s2-46.3.557}{Journal~of~the~London~Mathematical~Society
  \textbf{s2-46}, 557 (1992)\ignorespaces}\ignorespaces.

\bibitem{Graham-nonexistence}
C.~R. Graham,
\textit{``{Conformally Invariant Powers of the Laplacian, II: Nonexistence}''},
\doiref{10.1112/jlms/s2-46.3.566}{Journal~of~the~London~Mathematical~Society
  \textbf{s2-46}, 566 (1992)\ignorespaces}\ignorespaces.

\bibitem{Gover-nonexistence}
A.~R. {Gover} \& K.~{Hirachi},
\textit{``{Conformally invariant powers of the Laplacian -- A complete
  non-existence theorem}''},
\doiref{10.1090/S0894-0347-04-00450-3}{J.~Amer.~Math.~Soc. \textbf{17}, 389
  (2014)\ignorespaces}\ignorespaces.

\bibitem{Karananas:2015ioa}
G.~K. Karananas \& A.~Monin,
\textit{``{Weyl vs. Conformal}''},
\doiref{10.1016/j.physletb.2016.04.001}{Phys.~Lett.~B \textbf{757}, 257
  (2016)\ignorespaces}\ignorespaces,
\normalsize{\texttt{\arxivref{1510.08042}{arXiv:1510.08042
  \![hep-th]}}}\ignorespaces.

\bibitem{Farnsworth:2017tbz}
K.~Farnsworth, M.~A. Luty \& V.~Prilepina,
\textit{``{Weyl versus Conformal Invariance in Quantum Field Theory}''},
\doiref{10.1007/JHEP10(2017)170}{JHEP \textbf{1710}, 170
  (2017)\ignorespaces}\ignorespaces,
\normalsize{\texttt{\arxivref{1702.07079}{arXiv:1702.07079
  \![hep-th]}}}\ignorespaces.

\bibitem{Erdmenger:1997gy}
J.~Erdmenger,
\textit{``{Conformally covariant differential operators: Properties and
  applications}''},
\doiref{10.1088/0264-9381/14/8/008}{Class.~Quant.~Grav. \textbf{14}, 2061
  (1997)\ignorespaces}\ignorespaces,
\normalsize{\texttt{\arxivref{hep-th/9704108}{hep-th/9704108}}}\ignorespaces.

\bibitem{Erdmenger:1997wy}
J.~Erdmenger \& H.~Osborn,
\textit{``{Conformally covariant differential operators: Symmetric tensor
  fields}''},
\doiref{10.1088/0264-9381/15/2/003}{Class.~Quant.~Grav. \textbf{15}, 273
  (1998)\ignorespaces}\ignorespaces,
\normalsize{\texttt{\arxivref{gr-qc/9708040}{gr-qc/9708040}}}\ignorespaces.

\bibitem{Branson_1985}
T.~P. Branson,
\textit{``Differential operators cononically associated to a conformal
  structure.''},
\doiref{10.7146/math.scand.a-12120}{MATHEMATICA~SCANDINAVICA \textbf{57},
  293–345 (1985)\ignorespaces}\ignorespaces.

\bibitem{graham2005ambient}
C.~R. Graham \& K.~Hirachi,
\textit{``The ambient obstruction tensor and $Q$-curvature''},
in \textit{``AdS/CFT correspondence: Einstein metrics and their conformal
  boundaries''},
Eur. Math. Soc. Z{\"u}rich (2005)\ignorespaces,
\normalsize{\texttt{\arxivref{math/0405068}{math/0405068
  \![math.DG]}}}\ignorespaces.

\bibitem{Safari:2021ocb}
M.~Safari, A.~Stergiou, G.~P. Vacca \& O.~Zanusso,
\textit{``{Scale and conformal invariance in higher derivative shift symmetric
  theories}''},
\doiref{10.1007/JHEP02(2022)034}{JHEP \textbf{2202}, 034
  (2022)\ignorespaces}\ignorespaces,
\normalsize{\texttt{\arxivref{2112.01084}{arXiv:2112.01084
  \![hep-th]}}}\ignorespaces.

\bibitem{OsbornLectures}
H.~Osborn,
\textit{``{Lectures on Conformal Field Theories in more than two
  dimensions}''},
\href{http://www.damtp.cam.ac.uk/user/ho/CFTNotes.pdf}{\texttt{http://www.damtp.cam.ac.uk/user/ho/CFTNotes.pdf}}.

\bibitem{Osborn:2015rna}
H.~Osborn \& A.~Stergiou,
\textit{``{Structures on the Conformal Manifold in Six Dimensional
  Theories}''},
\doiref{10.1007/JHEP04(2015)157}{JHEP \textbf{1504}, 157
  (2015)\ignorespaces}\ignorespaces,
\normalsize{\texttt{\arxivref{1501.01308}{arXiv:1501.01308
  \![hep-th]}}}\ignorespaces.

\bibitem{FG1}
C.~Fefferman \& C.~R. Graham,
\textit{``Conformal invariants''},
in \textit{``\'Elie Cartan et les math\'ematiques d'aujourd'hui - Lyon, 25-29
  juin 1984''},
Soci\'et\'e math\'ematique de France (1985)\ignorespaces.

\bibitem{Osborn:1993cr}
H.~Osborn \& A.~C. Petkou,
\textit{``{Implications of conformal invariance in field theories for general
  dimensions}''},
\doiref{10.1006/aphy.1994.1045}{Annals~Phys. \textbf{231}, 311
  (1994)\ignorespaces}\ignorespaces,
\normalsize{\texttt{\arxivref{hep-th/9307010}{hep-th/9307010}}}\ignorespaces.

\bibitem{xact}
J.~Mart\'in-Garc\'ia,
\textit{``{xAct: Efficient Tensor Computer Algebra for Mathematica}''},
\href{http://www.xact.es}{\texttt{http://www.xact.es}}.

\bibitem{Brizuela:2008ra}
D.~Brizuela, J.~M. Martin-Garcia \& G.~A. Mena~Marugan,
\textit{``{xPert: Computer algebra for metric perturbation theory}''},
\doiref{10.1007/s10714-009-0773-2}{Gen.~Rel.~Grav. \textbf{41}, 2415
  (2009)\ignorespaces}\ignorespaces,
\normalsize{\texttt{\arxivref{0807.0824}{arXiv:0807.0824
  \![gr-qc]}}}\ignorespaces.

\bibitem{Nutma:2013zea}
T.~Nutma,
\textit{``{xTras : A field-theory inspired xAct package for mathematica}''},
\doiref{10.1016/j.cpc.2014.02.006}{Comput.~Phys.~Commun. \textbf{185}, 1719
  (2014)\ignorespaces}\ignorespaces,
\normalsize{\texttt{\arxivref{1308.3493}{arXiv:1308.3493
  \![cs.SC]}}}\ignorespaces.

\bibitem{Brown:1977pq}
L.~S. Brown \& J.~P. Cassidy,
\textit{``{Stress Tensor Trace Anomaly in a Gravitational Metric: General
  Theory, Maxwell Field}''},
\doiref{10.1103/PhysRevD.15.2810}{Phys.~Rev.~D \textbf{15}, 2810
  (1977)\ignorespaces}\ignorespaces.

\bibitem{Gibbons:2019lmj}
G.~W. Gibbons, C.~N. Pope \& S.~Solodukhin,
\textit{``{Higher Derivative Scalar Quantum Field Theory in Curved
  Spacetime}''},
\doiref{10.1103/PhysRevD.100.105008}{Phys.~Rev.~D \textbf{100}, 105008
  (2019)\ignorespaces}\ignorespaces,
\normalsize{\texttt{\arxivref{1907.03791}{arXiv:1907.03791
  \![hep-th]}}}\ignorespaces.

\bibitem{Penedones:2010ue}
J.~Penedones,
\textit{``{Writing CFT correlation functions as AdS scattering amplitudes}''},
\doiref{10.1007/JHEP03(2011)025}{JHEP \textbf{1103}, 025
  (2011)\ignorespaces}\ignorespaces,
\normalsize{\texttt{\arxivref{1011.1485}{arXiv:1011.1485
  \![hep-th]}}}\ignorespaces.

\bibitem{Fitzpatrick:2011dm}
A.~L. Fitzpatrick \& J.~Kaplan,
\textit{``{Unitarity and the Holographic S-Matrix}''},
\doiref{10.1007/JHEP10(2012)032}{JHEP \textbf{1210}, 032
  (2012)\ignorespaces}\ignorespaces,
\normalsize{\texttt{\arxivref{1112.4845}{arXiv:1112.4845
  \![hep-th]}}}\ignorespaces.

\bibitem{Bekaert:2015tva}
X.~Bekaert, J.~Erdmenger, D.~Ponomarev \& C.~Sleight,
\textit{``{Quartic AdS Interactions in Higher-Spin Gravity from Conformal Field
  Theory}''},
\doiref{10.1007/JHEP11(2015)149}{JHEP \textbf{1511}, 149
  (2015)\ignorespaces}\ignorespaces,
\normalsize{\texttt{\arxivref{1508.04292}{arXiv:1508.04292
  \![hep-th]}}}\ignorespaces.

\bibitem{Mikhailov:2002bp}
A.~Mikhailov,
\textit{``{Notes on higher spin symmetries}''},
\normalsize{\texttt{\arxivref{hep-th/0201019}{hep-th/0201019}}}\ignorespaces.

\bibitem{Braun:2003rp}
V.~M. Braun, G.~P. Korchemsky \& D.~M\"uller,
\textit{``{The Uses of conformal symmetry in QCD}''},
\doiref{10.1016/S0146-6410(03)90004-4}{Prog.~Part.~Nucl.~Phys. \textbf{51}, 311
  (2003)\ignorespaces}\ignorespaces,
\normalsize{\texttt{\arxivref{hep-ph/0306057}{hep-ph/0306057}}}\ignorespaces.

\bibitem{Ferrara:1972jwi}
S.~Ferrara, A.~F. Grillo \& R.~Gatto,
\textit{``{Logarithmic scaling and spontaneous breaking}''},
\doiref{10.1016/0370-2693(72)90077-9}{Phys.~Lett.~B \textbf{42}, 264
  (1972)\ignorespaces}\ignorespaces.

\bibitem{Gurarie:1993xq}
V.~Gurarie,
\textit{``{Logarithmic operators in conformal field theory}''},
\doiref{10.1016/0550-3213(93)90528-W}{Nucl.~Phys.~B \textbf{410}, 535
  (1993)\ignorespaces}\ignorespaces,
\normalsize{\texttt{\arxivref{hep-th/9303160}{hep-th/9303160}}}\ignorespaces.

\bibitem{Hofman:2008ar}
D.~M. Hofman \& J.~Maldacena,
\textit{``{Conformal collider physics: Energy and charge correlations}''},
\doiref{10.1088/1126-6708/2008/05/012}{JHEP \textbf{0805}, 012
  (2008)\ignorespaces}\ignorespaces,
\normalsize{\texttt{\arxivref{0803.1467}{arXiv:0803.1467
  \![hep-th]}}}\ignorespaces.

\bibitem{Bugini:2018def}
F.~Bugini \& D.~E. D\'\i{}az,
\textit{``{Holographic Weyl anomaly for GJMS operators: one Laplacian to rule
  them all}''},
\doiref{10.1007/JHEP02(2019)188}{JHEP \textbf{1902}, 188
  (2019)\ignorespaces}\ignorespaces,
\normalsize{\texttt{\arxivref{1811.10380}{arXiv:1811.10380
  \![hep-th]}}}\ignorespaces.

\bibitem{deBoer:2009pn}
J.~de~Boer, M.~Kulaxizi \& A.~Parnachev,
\textit{``{AdS(7)/CFT(6), Gauss-Bonnet Gravity, and Viscosity Bound}''},
\doiref{10.1007/JHEP03(2010)087}{JHEP \textbf{1003}, 087
  (2010)\ignorespaces}\ignorespaces,
\normalsize{\texttt{\arxivref{0910.5347}{arXiv:0910.5347
  \![hep-th]}}}\ignorespaces.

\bibitem{Bastianelli:2000hi}
F.~Bastianelli, S.~Frolov \& A.~A. Tseytlin,
\textit{``{Conformal anomaly of (2,0) tensor multiplet in six-dimensions and
  AdS / CFT correspondence}''},
\doiref{10.1088/1126-6708/2000/02/013}{JHEP \textbf{0002}, 013
  (2000)\ignorespaces}\ignorespaces,
\normalsize{\texttt{\arxivref{hep-th/0001041}{hep-th/0001041}}}\ignorespaces.

\bibitem{Codello:2019isr}
A.~Codello, M.~Safari, G.~P. Vacca \& O.~Zanusso,
\textit{``{Symmetry and universality of multifield interactions in
  $6-\varepsilon$ dimensions}''},
\doiref{10.1103/PhysRevD.101.065002}{Phys.~Rev.~D \textbf{101}, 065002
  (2020)\ignorespaces}\ignorespaces,
\normalsize{\texttt{\arxivref{1910.10009}{arXiv:1910.10009
  \![hep-th]}}}\ignorespaces.

\bibitem{Gies:2013noa}
H.~Gies \& S.~Lippoldt,
\textit{``{Fermions in gravity with local spin-base invariance}''},
\doiref{10.1103/PhysRevD.89.064040}{Phys.~Rev.~D \textbf{89}, 064040
  (2014)\ignorespaces}\ignorespaces,
\normalsize{\texttt{\arxivref{1310.2509}{arXiv:1310.2509
  \![hep-th]}}}\ignorespaces.

\bibitem{Lippoldt:2015cea}
S.~Lippoldt,
\textit{``{Spin-base invariance of Fermions in arbitrary dimensions}''},
\doiref{10.1103/PhysRevD.91.104006}{Phys.~Rev.~D \textbf{91}, 104006
  (2015)\ignorespaces}\ignorespaces,
\normalsize{\texttt{\arxivref{1502.05607}{arXiv:1502.05607
  \![hep-th]}}}\ignorespaces.

\bibitem{holland-sparling}
J.~Holland \& G.~Sparling,
\textit{``{Conformally invariant powers of the ambient Dirac operator}''},
\normalsize{\texttt{\arxivref{math/0112033}{math/0112033
  \![math.DG]}}}\ignorespaces.

\bibitem{Kubo:1977ye}
R.~Kubo,
\textit{``{Conformally Covariant Structure of the Dirac Equation}''},
\doiref{10.1143/PTP.58.2012}{Prog.~Theor.~Phys. \textbf{58}, 2012
  (1977)\ignorespaces}\ignorespaces.

\bibitem{Fischmann-2013}
M.~{Fischmann},
\textit{``{On Conformal Powers of the Dirac Operator on Spin Manifolds}''},
\normalsize{\texttt{\arxivref{1311.4182}{arXiv:1311.4182
  \![math.DG]}}}\ignorespaces.

\bibitem{deBerredo-Peixoto:2001uob}
G.~de~Berredo-Peixoto \& I.~L. Shapiro,
\textit{``{On the High derivative fermionic operator and trace anomaly}''},
\doiref{10.1016/S0370-2693(01)00801-2}{Phys.~Lett.~B \textbf{514}, 377
  (2001)\ignorespaces}\ignorespaces,
\normalsize{\texttt{\arxivref{hep-th/0101158}{hep-th/0101158}}}\ignorespaces.

\bibitem{amslaurea19852}
R.~Reho,
\textit{``A higher derivative fermion model''},
Laurea magistrale, Università di Bologna, Corso di Studio in
  Fisica\ignorespaces,
\href{http://amslaurea.unibo.it/19852/}{\texttt{http://amslaurea.unibo.it/19852/}}.

\bibitem{Anselmi:1999bu}
D.~Anselmi,
\textit{``{Irreversibility and higher spin conformal field theory}''},
\doiref{10.1088/0264-9381/17/15/301}{Class.~Quant.~Grav. \textbf{17}, 2847
  (2000)\ignorespaces}\ignorespaces,
\normalsize{\texttt{\arxivref{hep-th/9912122}{hep-th/9912122}}}\ignorespaces.

\bibitem{Fischmann2013}
M.~Fischmann,
\textit{``Conformally covariant differential operators acting on spinor bundles
  and related conformal covariants''},
Doctoral dissertation, Humboldt-Universität zu Berlin,
  Mathematisch-Naturwissenschaftliche Fakultät II\ignorespaces,
\href{http://dx.doi.org/10.18452/16703}{\texttt{http://dx.doi.org/10.18452/16703}}.

\bibitem{Frob:2020gdh}
M.~B. Fr\"ob,
\textit{``{FieldsX -- An extension package for the xAct tensor computer algebra
  suite to include fermions, gauge fields and BRST cohomology}''},
\normalsize{\texttt{\arxivref{2008.12422}{arXiv:2008.12422
  \![hep-th]}}}\ignorespaces.

\bibitem{Gliozzi:2016ysv}
F.~Gliozzi, A.~Guerrieri, A.~C. Petkou \& C.~Wen,
\textit{``{Generalized Wilson-Fisher Critical Points from the Conformal
  Operator Product Expansion}''},
\doiref{10.1103/PhysRevLett.118.061601}{Phys.~Rev.~Lett. \textbf{118}, 061601
  (2017)\ignorespaces}\ignorespaces,
\normalsize{\texttt{\arxivref{1611.10344}{arXiv:1611.10344
  \![hep-th]}}}\ignorespaces.

\bibitem{Gracey:2017erc}
J.~A. Gracey \& R.~M. Simms,
\textit{``{Higher dimensional higher derivative $\phi^4$ theory}''},
\doiref{10.1103/PhysRevD.96.025022}{Phys.~Rev.~D \textbf{96}, 025022
  (2017)\ignorespaces}\ignorespaces,
\normalsize{\texttt{\arxivref{1705.06983}{arXiv:1705.06983
  \![hep-th]}}}\ignorespaces.

\bibitem{Safari:2017irw}
M.~Safari \& G.~P. Vacca,
\textit{``{Multicritical scalar theories with higher-derivative kinetic terms:
  A perturbative RG approach with the $\varepsilon$-expansion}''},
\doiref{10.1103/PhysRevD.97.041701}{Phys.~Rev.~D \textbf{97}, 041701
  (2018)\ignorespaces}\ignorespaces,
\normalsize{\texttt{\arxivref{1708.09795}{arXiv:1708.09795
  \![hep-th]}}}\ignorespaces.

\bibitem{Safari:2017tgs}
M.~Safari \& G.~P. Vacca,
\textit{``{Uncovering novel phase structures in $\Box ^k$ scalar theories with
  the renormalization group}''},
\doiref{10.1140/epjc/s10052-018-5721-4}{Eur.~Phys.~J.~C \textbf{78}, 251
  (2018)\ignorespaces}\ignorespaces,
\normalsize{\texttt{\arxivref{1711.08685}{arXiv:1711.08685
  \![hep-th]}}}\ignorespaces.

\end{thebibliography}

\end{document}